%% file: pscc2024_paper.tex
\documentclass{IEEEtran4PSCC}
\ifCLASSINFOpdf
   \usepackage[pdftex]{graphicx}
\else
   \usepackage[dvips]{graphicx}
\fi
\usepackage{cuted} 
\usepackage[cmex10]{amsmath}
\usepackage{amsfonts}
\usepackage{amssymb}
\usepackage{amsthm}

\usepackage{accents}
\newcommand{\ubar}[1]{\underaccent{\bar}{#1}}

\usepackage{float}
\ifCLASSOPTIONcompsoc
 \usepackage[caption=false,font=normalsize,labelfont=sf,textfont=sf]{subfig}
\else
 \usepackage[caption=false,font=footnotesize]{subfig}
\fi

\floatstyle{ruled}
\newfloat{model}{thp}{lop}
\floatname{model}{Model}

\usepackage{booktabs}

\usepackage{todonotes}

\usepackage{enumitem}
\usepackage{hyperref}
\usepackage{tikz-3dplot}
\usepackage{multirow}
\newlist{abbrv}{itemize}{1}
\setlist[abbrv,1]{label=,labelwidth=0.5in,align=parleft,itemsep=0.1\baselineskip,leftmargin=!}

\begin{document}
%
\title{High-Spatial Resolution Transmission and Storage Expansion Planning for High Renewable Grids: \\ A Case Study}


\author{\IEEEauthorblockN{Kevin Wu\IEEEauthorrefmark{1},
Rabab Haider\IEEEauthorrefmark{2},
Pascal Van Hentenryck\IEEEauthorrefmark{1}}
\IEEEauthorblockA{\IEEEauthorrefmark{1} Georgia Institute of Technology, Atlanta, GA, United States\\ 
kwu381@gatech.edu, pvh@gatech.edu}
\IEEEauthorblockA{\IEEEauthorrefmark{2} University of Michigan, Ann Arbor, MI, United States\\
rababh@umich.edu}
}

\maketitle

\begin{abstract}
    Transmission Expansion Planning (TEP) is the process of optimizing the development and upgrade of the power grid to ensure reliable, efficient, and cost-effective electricity delivery while addressing grid constraints. To support growing demand and renewable energy integration, energy storage is emerging as a pivotal asset that provides temporal flexibility and alleviates congestion. This paper presents a TEP model that incorporates the sizing and siting of short-duration storage. With a focus on high spatial resolution, the model is applied to a 2,000-bus synthetic Texas power system, offering detailed insights into geographic investment and operational patterns. To maintain computational feasibility, a simple yet effective storage candidates (SC) method is introduced, significantly reducing the search space. Results highlight that transmission investments are primarily driven by renewable energy expansion, while storage investments are shaped by renewable curtailment and load-shedding events, with their primary function being peak load shaving. The findings underscore the importance of co-optimizing transmission and storage to minimize costs and enhance grid reliability. However, limitations in the ability of the SC method to identify optimal storage locations to meet long-term needs suggest opportunities for future research, including dynamic candidate selection and hybrid optimization techniques.
\end{abstract}


\thanksto{\noindent This research was partly supported by the NSF AI Institute for Advances in Optimization (Award No. 2112533), and NSF Global Center Electric Power Innovation for a Carbon-free Society (Award No. 2330450).}

\input{nomenclature}

\input{introduction}

\input{formulation}

\input{methodology}

\input{casestudy}

\input{results}

\input{conclusion}

\bibliographystyle{IEEEtran}
\bibliography{refs}

\end{document}

%% file: nomenclature.tex
\section*{Nomenclature}
\label{sec:nomenclature}

\newcommand{\db}{\delta B}
\newcommand{\dbhi}{\bar{\delta B}}
\newcommand{\dblo}{\ubar{\delta B}}
\newcommand{\dva}{{\Delta \theta}}
\newcommand{\dvahi}{\bar{\Delta \theta}}
\newcommand{\dvalo}{\ubar{\Delta \theta}}

\paragraph{Sets}

\renewcommand{\S}{\mathcal{S}}
\newcommand{\N}{\mathcal{N}}
\newcommand{\T}{\mathcal{T}}
\newcommand{\E}{\mathcal{E}}

\begin{abbrv}
    \item[$\N$] Set of buses; $\N = \{1, ..., N\}$; indexed by $i$
    \item[$\E$] Set of branches; $\E = \{1, ..., E\}$; indexed by $ij$
    \item[$\S$] Set of scenarios; $\S = \{1, ..., S\}$; indexed by $s$
    \item[$\T$] Set of hours; $\T = \{1, ..., T\}$; indexed by $t$
\end{abbrv}
\vspace{\baselineskip} 
Note: Variables and parameters are indexed by $i$ (bus), $s$ (scenario), and $t$ (hour) where applicable. If indices are not included, they refer to node $i$, in scenario $s$, and at hour $t$.

\paragraph{Parameters}

\newcommand{\pd}{\mathbf{p}^{\text{d}}}
\newcommand{\pgmin}{\mathbf{\ubar{p}}^{\text{g}}}
\newcommand{\pgmax}{\mathbf{\bar{p}}^{\text{g}}}
\newcommand{\pfmax}{\mathbf{\bar{p}}^{\text{f}}}
\newcommand{\dvamin}{\ubar{\theta}}
\newcommand{\dvamax}{\bar{\theta}}

\newcommand{\df}{\delta \pf}
\newcommand{\dX}{\delta X}
\newcommand{\dXmin}{\ubar{\delta X}}
\newcommand{\dXmax}{\bar{\delta X}}
\newcommand{\dB}{\delta B}
\newcommand{\dBmin}{\ubar{\delta B}}
\newcommand{\dBmax}{\bar{\delta B}}

\begin{abbrv}
    \item[$\pd_{i,s,t}$] Load at node $i$ in scenario $s$ at hour $t$
    \item[$\pgmin_{i,s,t}$] min output of generator $i$ in scenario $s$ at hour $t$
    \item[$\pgmax_{i,s,t}$] max output of generator $i$ in scenario $s$ at hour $t$
    \item[$c^{\text{cap}}_{ij}$] Line capacity upgrade cost of edge $ij$
    \item[$c^{\text{s-cap}}_{i}$] Storage fixed installation cost at node $i$
    \item[$c^{\text{PR}}_{i}$] Storage power rating cost at node $i$
    \item[$c^{\text{ER}}_{i}$] Storage energy rating cost at node $i$
    \item[$\bar{P}^{\text{storage}}$] max storage power rating
    \item[$\bar{E}^{\text{storage}}$] max storage energy rating
    \item[$c^{\text{G}}_{i}(\cdot)$] Production cost function of generator $i$
    \item[$\Omega_{\S}(\cdot)$] Probability mass function for set of scenarios $\S$
    \item[$\lambda$] Power balance violation penalty cost
    \item[$\Delta^c_{ij}$] Capacity upgrade increment (MW) of branch $ij$
    \item[$m$] Integer number of capacity upgrade increments available
    \item[$X_{ij}$] Reactance of branch $ij$
    \item[$\pfmax_{ij}$] Thermal limit of branch $ij$
    \item[$\dvamin_{ij}, \dvamax_{ij}$] min/max angle difference on branch $ij$
\end{abbrv}

\paragraph{Variables}

\newcommand{\pg}{\mathbf{p}^{\text{g}}}
\newcommand{\pf}{\mathbf{p}^{\text{f}}}
\newcommand{\va}{\theta}

\begin{abbrv}
    \item[$\gamma_{ij}$] Integer capacity upgrade level of branch $ij$
    \item[$\sigma_{i}$] Binary indicating whether energy storage is installed on node $i$
    \item[$s^{\text{PR}}_i$] Power rating of storage at node $i$
    \item[$s^{\text{ER}}_i$] Energy rating of storage at node $i$
    \item[$s_{i,s,t}$] State of charge of storage at node $i$ in scenario $s$ at hour $t$
    \item[$\text{ch}_{i,s,t}$] Amount to charge storage at node $i$ in scenario $s$ at hour $t$
    \item[$\text{dis}_{i,s,t}$] Amount to discharge storage at node $i$ in scenario $s$ at hour $t$
    \item[$\va_{i, s, t}$] Voltage angle of bus $i$ in scenario $s$ at hour $t$
    \item[$\pg_{i,s,t}$] Output of generator $i$ in scenario $s$ at hour $t$
    \item[$\pf_{ij,s,t}$] Power flow on branch $ij$ in scenario $s$ at hour $t$
    \item[$\xi_{i,s,t}$] Power imbalance at bus $i$ in scenario $s$ at hour $t$
    \item[$\alpha_{i,s,t}$] Storage linearizing binary variable at node $i$ in scenario $s$ at hour $t$ 
    \item[$\beta_{i,s,t}$] Storage linearizing (continuous) variable at node $i$ in scenario $s$ at hour $t$
\end{abbrv}

%% file: introduction.tex
\section{Introduction}
\label{sec:introduction}

The U.S. electric grid is undergoing a significant transformation, with the increased penetration of renewable energy resources such as wind and solar. Combined with the anticipated growth in electricity demand, these shifts necessitate substantial investments in transmission infrastructure to maintain reliable and affordable access to electricity \cite{crossedwires2024}. This presents a critical challenge for transmission system operators, who are tasked with ensuring the grid is capable of delivering power from generators to load centers. Addressing this challenge requires solving Transmission Expansion Planning (TEP) problems, which focus on identifying cost-effective strategies for grid expansion, while accounting for long-term changes in generation capacity and demand patterns \cite{Cho}. 

The TEP problem is a mixed-integer program (MIP) where integer variables represent the yes/no investment decisions on upgrading existing lines or building new lines. It is computationally challenging to solve due to the presence of integer variables in large-scale models that represent the long time-frame of investment decisions and various grid conditions. Further extensions to the TEP problem include additional investments decisions, such as generation in joint generation-transmission expansion models \cite{hemmati2013comprehensive}, FACTS devices \cite{Wu_2024_TNEPFACTS}, or energy storage \cite{Sheibani_review}. Notably, energy storage provides necessary operational flexibility in modern power systems with high penetrations of renewable resources \cite{Denholm_2011_storageflexibility}. In the U.S., approximately 16 GW of storage capacity is planned or operational as of 2023, with this number expected to double by the end of 2024 \cite{EIA2024BatteryStorage}. Storage can provide critical benefits, including congestion reduction, peak shaving, and mitigating the uncertainties associated with variable renewable generation \cite{behabtu2020review}. Motivated by these observations, this paper studies joint TEP+Storage investments to enhance grid reliability in future high-renewable generation scenarios, while minimizing the cost of investment and operations. Prior studies of the TEP+Storage problem predominantly focus on aggregated models with low spatial resolution, limiting their applicability to real-world scenarios. The focus of this paper is thus on improving the computational tractability of TEP+Storage investment decisions at a high spatial resolution, and to use the resulting model to understand future investment decisions, and the specific roles of transmission and storage expansions.

\subsection{Effects of Model Resolution on Planning}
\label{sec:introduction:resolutions}
The energy system models used for planning decisions can be described along the three axis of operational, temporal, and spatial resolution. These are reviewed below, and the case for higher spatial resolutions is presented.

\textbf{Operational resolution} pertains to the power system's governing rules, particularly the power flow model used \cite{Jacobson}. The simplest model is a transport flow, which only models nodal power balance to ensure flow conservation. This simple model provides useful information for the overall transfer capacity between regions, but is unable to capture transmission bottlenecks in models with high spatial resolution \cite{Neumann_2022_flowmodelsforCEP}. The transport model can be extended to a linearized DC model (and its multiple reformulations), which includes additional linearized constraints for generalized Ohm's law in Kirchhoff’s Voltage Law. The most detailed models are AC models which are nonlinear and nonconvex, but may be convexified or linearized (as with the DC approximation). Most TEP+Storage studies adopt the DC approximation, as energy storage primarily affects real power, with limited impact on reactive power \cite{Xiong}. Additional operational constraints include ramping constraints on generators, system reserve requirements, and unit commitment (UC) decisions for generators. Using higher operational resolutions introduces additional system constraints, additional variables (i.e., binary commitment decisions for UC constraints), and inter-temporal coupling constraints (i.e., ramping, reserves, UC, storage operations).

\textbf{Temporal resolution} refers to time-dependent aspects, such as load and renewable generation, and is typically modeled at an hourly resolution, referred to as the subperiod. Year-long operations are often modeled using representative periods, typically 24-hour durations, to reduce the computational time required for evaluating a full year of operations. Some studies use finer time steps (e.g., 10-minute subperiods) to capture system ramping constraints \cite{Gan}, while others employ representative periods of varying durations to capture weather patterns, weekly demand, or the operations of certain committed generators or long-duration storage \cite{JacobsonPecci, Piansky}. The interaction between the length of each representative period (e.g., 24 hours) and the subperiod’s operational resolution (e.g., hourly or 10-minute intervals) is critical, as capturing fast system responses requires sub-hourly subperiods. However, such fine temporal resolution becomes computationally prohibitive for long representative periods, such as a week, which contains 1,008 10-minute subperiods coupled with inter-temporal constraints.

\textbf{Spatial resolution} describes the granularity of the network, such as the number of buses or zones. Many studies aggregate buses into zones to simplify spatial complexity. For instance, the NREL ReEDS model \cite{ReEDS} aggregates the contiguous U.S. into 134 zones and incorporates emissions and reserves constraints. In contrast, the GenX model used in \cite{JacobsonPecci} aggregates into 26 zones, handles UC and ramping constraints, and employs representative periods of 168 hours. To ensure computational tractability, researchers often reduce resolution in one or more dimensions, which can impact siting, costs, and operations \cite{Jacobson, XuHobbs}. Overall model accuracy is limited by the lowest-resolution dimension, highlighting the need for computational improvements jointly across all resolutions. A recent study further demonstrates that the \textit{errors in investment decisions are greatest for models with low spatial resolution}, as they lack the locational accuracy to sufficiently inform decision making \cite{Jacobson}. 

\emph{This paper addresses this limitation by presenting a new, simple yet effective, search space reduction method to ensure computational tractability of storage investment decisions. The method can then be applied to a case study of TEP+Storage on the U.S. Texas grid at a high spatial resolution, providing realistic insights for future transmission and storage investments.}

\subsection{Related Work}
\label{sec:introduction:literature}
The TEP+Storage problem is the focus of considerable research, due to the growing importance of transmission and storage in integrating renewable energy and enhancing grid reliability. Numerous studies explore its formulation and solution techniques, typically modeled as a MIP with additional decision variables for storage sizing, siting, and operations \cite{Sheibani_review}.

Table~\ref{tab:literature_overview} provides an overview of TEP+Storage approaches which model the investment decisions as an MIP. Several prior work co-optimize generation expansion planning (GEP), transmission expansion planning (TEP), and energy storage system (ESS) investments (e.g., \cite{ReEDS, JacobsonPecci, Ansari, Moradi, Gan, Dvorkin, MacRae}), while others consider only ESS investments (e.g., \cite{Xiong, Blanco, Piansky}). The table also provides details on whether the study considers line upgrades or new construction, the type of power model used, the lengths of subperiods and representative periods, and the spatial resolution of each study.

\begin{table*}[t]
\centering
\caption{Overview of TEP+Storage literature}
\label{tab:literature_overview}
\begin{tabular}{lllllll}
\toprule
 & \textbf{Co-Investments} & \textbf{Lines} & \textbf{Power model} & \textbf{Subperiod} & \textbf{Rep. period} & \textbf{Spatial} \\
\midrule
Ho et al. 2021 \cite{ReEDS} & GEP, TEP, ESS & upgrade & transport & hour & day & 134-zone \\
Jacobson et al. 2024 \cite{JacobsonPecci} & GEP, TEP, ESS & upgrade & transport & hour & week & 26-zone \\
Ansari et al. 2021 \cite{Ansari} & GEP, TEP, ESS & new & DC, Linearized AC & hour & day & 6, 24-bus \\
Moradi-Sepahvand et al. 2021 \cite{Moradi} & GEP, TEP, ESS & new & DC & hour & 96 hours & 24-bus \\
Gan et al. 2019 \cite{Gan} & TEP, ESS & new & DC & 5 min. & day & 6, 53-bus \\
Dvorkin et al. 2017 \cite{Dvorkin} & TEP, ESS & new & DC & hour & day & 240-bus \\
MacRae et al. 2016 \cite{MacRae} & TEP, ESS & new & DC & 30min. - hour & day & 6, 25, 46-bus \\
Xiong et al. 2015 \cite{Xiong} & ESS & - & DC & hour & day & 24-bus \\
Blanco et al. 2016 \cite{Blanco} & ESS & - & DC & hour & day & 240-bus \\
Piansky et al. 2024 \cite{Piansky} & ESS & - & DC & hour & week & 240-bus \\
\textbf{This paper} & TEP, ESS & upgrade & DC & hour & day & \textbf{2000-bus} \\
\bottomrule
\end{tabular}
\end{table*}


The studies at the highest operational resolution focus on more accurate power flow models and additional operational constraints \cite{Ansari, Moradi}. In \cite{Moradi} the focus is on modeling forward ramping spinning reserves and evaluating low-carbon policy implications. In \cite{Ansari} the adaptability of power systems to uncertainty is evaluated using a flexibility index. The authors propose a stochastic program using linearized AC constraints via first-order Taylor series and big-M methods. 

The studies at the highest temporal and operational resolution consider sub-hourly operations and a security-constrained DC model \cite{Gan}. They assess the use of batteries for corrective control with a 5-minute resolution, using a security-constrained DC model. Generation uncertainty is also considered in \cite{Xiong}, which uses scenario trees to capture the uncertainty in wind generation in storage investment decisions. 

To achieve computational tractability, Benders decomposition is frequently used \cite{Gan, Xiong, MacRae, JacobsonPecci}. The storage operations are simplified in \cite{MacRae} by assuming perfect storage efficiency; this linearizes the operation of storage and permits the use of the Benders approach. All of these studies evaluate their methods at low spatial resolutions, the largest being a 53-bus system.

The studies at the highest spatial resolution utilize a 240-bus model of the Western Electricity Coordinating Council (WECC) \cite{Dvorkin, Blanco, Piansky}, which covers 11 U.S. states and the Canadian provinces of British Columbia and Alberta. A tri-level optimization program is developed in \cite{Dvorkin} to solve the co-planning of transmission infrastructure and merchant storage. The first level maximizes storage operator profits, the second level conducts transmission expansion planning, and the third level performs market-clearing. In this work, the authors consider an additional resolution of energy system stakeholders -- both privately-owned merchant storage and transmission system operators are modelled. In \cite{Piansky} both high spatial and high temporal resolution are considered. The model evaluates operations over week-long representative periods at hourly operating resolution, to investigate long-duration storage under public safety power shutoff scenarios for wildfire risk mitigation. Finally, \cite{Blanco} consider both high spatial and operational resolution. They introduce a linearization technique to model storage operation while preventing simultaneous charge and discharge by introducing additional binary variables to the operations.


\subsection{Contributions}
\label{sec:introduction:contribution}

The main contribution of the paper is to present a realistic methodology for co-optimizing TEP and storage investments at a spatial resolution of 2000 buses, achieving a higher level of detail than test cases in prior literature. This level of fidelity makes it possible to present a significant case study and identify the key roles of transmission and storage investment options. More precisely, the contributions of the paper can be summarized as follows.

\begin{itemize}
    \item The paper presents a novel, simple but effective storage candidates heuristic that makes the TEP problem with storage  computationally feasible for realistic grids. 
    \item Co-investment in line upgrades and storage investments is necessary to meet electricity demand of future years while keeping costs low. Neither technology alone can meet future load.
    \item Transmission is the primary investment pathway to en-able high utilization of zero-cost renewable energy by reducing congestion in critical regions. Storage is the primary investment pathway to satisfy peak demand in the afternoon via peak shaving and load shifting actions.
\end{itemize}

\noindent
The rest of the paper is organized as follows. Section \ref{sec:formulation} describes the TEP+Storage formulation. Section \ref{sec:methodology} presents the proposed search space reduction to identify candidate storage investment nodes, and Section \ref{sec:casestudy} presents the case study of the Texas grid. Section \ref{sec:results} presents the results of TEP+Storage, highlighting the computational advantages of the proposed formulation and the investment decisions necessary to reliably meet projected electricity demand. Section \ref{sec:conclusion} concludes the paper and highlights directions for future work.

%% file: formulation.tex
\section{Formulation}
\label{sec:formulation}

This section presents the TEP+Storage formulation used in the paper. It considers a single generator connected to each bus, without loss of generality. Reference (slack) bus voltage constraints are also omitted for simplicity.

\subsection{The TEP+Storage Formulation}
\label{sec:formulation:TNEP}

    The TEP+Storage model is structured as a two-stage formulation, consisting of an investment problem and a recourse problem. The investment stage determines capacity upgrades for transmission lines and the sizing and siting of energy storage installations, minimizing capital expenditures ($CapEx$). The recourse stage evaluates the operational feasibility and costs ($OpEx$) over representative days, considering hourly generation dispatch and load-shedding penalties. This structure captures the interplay between long-term planning decisions and short-term operational dynamics.

    Model \ref{model:TNEP} outlines the TEP+Storage formulation, with the following sections detailing its variables, constraints, and objective function.
    
    \begin{model}[!t]
        \caption{TEP+Storage Formulation}
        \label{model:TNEP}
        \begin{align*}
            \text{min} \quad 
                & CapEx
                + OpEx
                \\
            \text{s.t.} \quad 
            & \eqref{eq:TNEP:power_balance}-\eqref{eq:TNEP:generation:min_max_limits},
                \quad \forall i \in \N, s \in \S, t \in \T \\
            & \eqref{eq:TNEP:ohm}-\eqref{eq:TNEP:phase_angle_difference},
                \quad \forall ij \in \E, s \in \S, t \in \T \\ 
            & \eqref{eq:TNEP:state_of_charge}
                \quad \forall i \in \N, s \in \S, t \in \{2,...,T-1\} \\
            & \eqref{eq:TNEP:initial_charge}-\eqref{eq:TNEP:final_charge},
                \quad \forall i \in \N, s \in \S \\
            & \eqref{eq:TNEP:soc_energy_rating}
            \quad \forall i \in \N, s \in \S, t \in \T \\
            & \eqref{eq:TNEP:storage_installed_energy} - \eqref{eq:TNEP:short_duration},
            \quad \forall i \in \N \\
            & \eqref{eq:TNEP:storage_linearizing_constraints_1} - \eqref{eq:TNEP:storage_linearizing_constraints_4},
            \quad \forall i \in \N, s \in \S, t \in \T
        \end{align*}
    \end{model}

    \subsubsection{Investment decisions and variables}
    \label{sec:formulation:TNEP:variables}

    In this study, \emph{capacity upgrades for existing transmission lines} are prioritized, as constructing new lines is challenging due to various economic, political, and social constraints. Upgrades to existing lines can take 18 months to 3 years, compared to 10 or more years for building new transmission lines, and have significantly lower costs \cite{GridLab2024Reconductoring, TANC_TransmissionQandA}. These capacity upgrades are modeled using investment variables $\gamma_{ij}$, which represent the upgrade level for each branch $ij \in \mathcal{E}$. The variables are restricted to integer values, allowing for up to $m$ distinct levels of capacity enhancement per line. 
    
    \emph{Short-duration lithium-ion batteries} are considered for energy storage. They make up the majority of existing utility scale storage devices due to their high energy and power density, high roundtrip efficiency, and high technology readiness level (TLR of 9 out of 9) \cite{MIT2022FutureEnergyStorage}. Storage investments are modeled using a binary variable $\sigma_{i}$, which indicates the decision to install storage at a node $i \in \mathcal{N}$. The power and energy ratings of the storage unit are represented by continuous variables $s^{PR}_i$ and $s^{ER}_i$, respectively. The operational dynamics of the storage at each hour $t \in \mathcal{T}$ are captured through variables $\text{ch}_{i,t}$, $\text{dis}_{i,t}$, and $s_{i,t}$, representing the energy charged, energy discharged, and state of charge at node $i$. The storage model incorporates the linearization formulation introduced in \cite{Blanco}, enabling accurate modeling of storage efficiency while preventing simultaneous charging and discharging. This formulation introduces additional variables: a binary variable $\alpha_{i,t}$ and a continuous variable $\beta_{i,t}$ (see constraints \eqref{eq:TNEP:storage_linearizing_constraints_1} -\eqref{eq:TNEP:storage_linearizing_constraints_4} in Section \ref{sec:formulation:TNEP:constraints} for further details).
    
    The model also incorporates other key variables, including nodal generation dispatch $\pg$, nodal voltage angles $\va$, power flows $\pf$, and nodal power imbalance variables $\xi$.

    \subsubsection{Constraints}
        \label{sec:formulation:TNEP:constraints}
        
        The power balance equation is enforced at each bus $i \in \mathcal{N}$, for every scenario $s \in \mathcal{S}$, and each hour $t \in \mathcal{T}$ as follows:
        \begin{equation}
        \label{eq:TNEP:power_balance}
        \begin{split}
            \pg_{i,s,t} + \sum_{ji \in \mathcal{E}} \pf_{ji,s,t} - \sum_{ij \in \mathcal{E}} \pf_{ij,s,t} + \text{dis}_{i,s,t} = \\ \, \pd_{i,s,t} 
            + \text{ch}_{i,s,t} + \xi_{i,s,t},
        \end{split}
        \end{equation}
        where the slack variable $\xi_{i,s,t}$ accounts for any power imbalances at the node. Any energy imbalance is penalized in the objective function.
        
        The output of each generator is bounded by its minimum and maximum capacity limits:
        \begin{align}
        \label{eq:TNEP:generation:min_max_limits}
            \pgmin_{i,s,t} \leq \pg_{i,s,t} \leq \pgmax_{i,s,t}.
        \end{align}
        Renewable generators, such as wind and solar, have a minimum output of zero, with maximum generation levels that vary according to the scenario $s$. For non-renewable generators, the capacity limits are typically based on physical attributes and may differ between seasonal periods (e.g., summer vs. winter). Non-renewable generators are treated as dispatchable generating units. Renewable generators can be curtailed.
        
        The power flow on each branch $ij \in \mathcal{E}$ is determined using Ohm's law:
        \begin{align}
        \label{eq:TNEP:ohm}
            \pf_{ij,s,t} = \frac{1}{X_{ij}} (\theta_{j,s,t} - \theta_{i,s,t}),
        \end{align}
        and must adhere to the following thermal limit constraints:
        \begin{align}
        \label{eq:TNEP:thermal_limit}
            -\pfmax_{ij} - \gamma_{ij} \Delta^{c}_{ij} \leq \pf_{ij,s,t} \leq \pfmax_{ij} + \gamma_{ij} \Delta^{c}_{ij}.
        \end{align}
        The thermal limit $\pfmax_{ij}$ can be increased through line upgrades, as represented by the investment variable $\gamma_{ij}$. The formulation does not model the corresponding change in branch reactance when reconductoring a line selected for upgrades. Additionally, the difference in phase angles, $\va_{ij,s,t} = (\va_{j,s,t} - \va_{i,s,t})$, is restricted by the following bounds:
        \begin{align}
        \label{eq:TNEP:phase_angle_difference}
            \dvamin_{ij} \leq \va_{ij,s,t} \leq \dvamax_{ij}.
        \end{align}
        The following constraints ensure the proper temporal evolution of the state of charge (SoC) for each storage unit. They also enforce a half-charge condition at the start and end of each representative day.
        \begin{align}
            \label{eq:TNEP:state_of_charge}
            s_{i,s,t} = s_{i,s,t-1} + \text{ch}_{i,s,t}\eta - \text{dis}_{i,s,t}/\eta
        \end{align}
        The constraint above applies for all hours except the first and last, i.e., $t \in \{2,...,T-1\}$. The initial and final charge constraints are given by
        \begin{align}
            \label{eq:TNEP:initial_charge}
            s_{i,s,1} = \frac{1}{2}s^\text{ER}_{i} + \text{ch}_{i,s,1}\eta - \frac{\text{dis}_{i,s,1}}{\eta}
        \end{align}
        \begin{align}
            \label{eq:TNEP:final_charge}
            s_{i,s,T} = \frac{1}{2}s^\text{ER}_{i}
        \end{align}
        The SoC at each storage node is constrained by its energy rating:
        \begin{align}
            \label{eq:TNEP:soc_energy_rating}
            0 \leq s_{i,s,t} \leq s^\text{ER}_{i}
        \end{align}
        Similarly, the energy and power ratings of each storage unit depend on whether storage is installed at the node, as represented by the investment variable $\sigma_i$ and are bounded by the specified parameters:
        \begin{align}
            \label{eq:TNEP:storage_installed_energy}
            0 \leq s^\text{ER}_{i} \leq \sigma_i \bar{E}^{\text{storage}}
        \end{align}
        \begin{align}
            \label{eq:TNEP:storage_installed_power}
            0 \leq s^\text{PR}_{i} \leq \sigma_i \bar{P}^{\text{storage}}
        \end{align}
        To account for the physical limitations of lithium-ion batteries as a short-duration resource, an additional constraint restricts the duration to a maximum of 4 hours.
        \begin{align}
            \label{eq:TNEP:short_duration}
            s^{ER}_{i} \leq 4 s^{PR}_{i}
        \end{align}
        The following constraints enforce the charge and discharge rates to lie within the power rating while also preventing simultaneous charging and discharging at each storage node. The linearization technique of \cite{Blanco} is adapted.
        \begin{align}
            0 & \leq \eta \, \text{ch}_{i,r,t} \leq \beta_{i,r,t}, \label{eq:TNEP:storage_linearizing_constraints_1}\\
            0 & \leq \frac{1}{\eta} \text{dis}_{i,r,t}  \leq s^{PR}_{i} - \beta_{i,r,t}, \label{eq:TNEP:storage_linearizing_constraints_2} \\
            0 & \leq \beta_{i,r,t} \leq \alpha_i \bar{P}^{\text{storage}}, \label{eq:TNEP:storage_linearizing_constraints_3} \\
            0 & \leq s^{PR}_{i} - \beta_{i,r,t} \leq (1 - \alpha_{i,r,t}) \bar{P}^{\text{storage}}, \label{eq:TNEP:storage_linearizing_constraints_4} 
        \end{align}

    \subsubsection{Objective}
    \label{sec:formulation:TNEP:objective}
    
    The objective of the TEP+Storage model is to minimize the total costs of investments for each period, and operating costs for each representative day. The investment costs, denoted as capital expenditures $CapEx$, include the  investment in line capacity upgrades and storage installations. The operational costs, denoted as operational expenditures $OpEx$, include the generation costs paid to the generators, $GenEx$, and any penalties incurred for unserved demand occurring during a load shedding event. The $OpEx$ is calculated over an entire year of operations, while the $CapEx$ is taken as a single cost over the particular investment period.

    \begin{align*}
        CapEx := \sum_{ij \in \E} c^{\text{cap}}_{ij} \gamma_{ij} + \sum_{i \in \N}{ c^{\text{s-cap}}_i\sigma_i + c^{\text{PR}}_is^\text{PR}_i + c^\text{ER}_is^\text{ER}_i}
    \end{align*}

    \begin{align*}
        OpEx &:= GenEx + Penalties \\
        &\;= 365 \sum_{i \in \N, s \in \S, t \in \T} 
                    \Omega_{\S}(s) [ c^\text{G}_{i}(\pg_{i,s,t})
                    + \lambda |\xi_{i,s,t}| ]
    \end{align*}

%% file: methodology.tex
\section{Methodology}
\label{sec:methodology}

A key focus of this paper is to improve the computational tractability of TEP+Storage investment decisions at a high spatial resolution. Preliminary simulation results indicate the storage investment decisions are the computational bottleneck. The following sections present a simple but effective search space reduction method for candidate nodes. The proposed method is evaluated on a case study of the Texas system, with a spatial resolution of 2000 buses (see Section~\ref{sec:casestudy}).

\subsection{Search Space Reduction via Storage Candidates (SC)}

Modeling storage operations in the recourse problem which accounting for storage efficiency introduces significant complexity in Model~\ref{model:TNEP}. Preventing simultaneous charging and discharging of storage devices requires either binary variables \cite{Blanco} or nonconvexities \cite{Yildiran}. Model \ref{model:TNEP} employs the binary variable $\alpha_{i,r,t}$ for the storage recourse. With 2000 buses under consideration for storage and 24-hour representative periods, this results in 48,000 binary variables solely for storage operations in the recourse problem, in addition to the 2000 binaries for the original storage investment siting decision. This rapidly escalates the problem’s computational complexity, making it computationally intractable at high spatial resolutions, even with access to high performance computing clusters.

\begin{figure}[!t]
    \centering
    \includegraphics[width=0.8\linewidth]{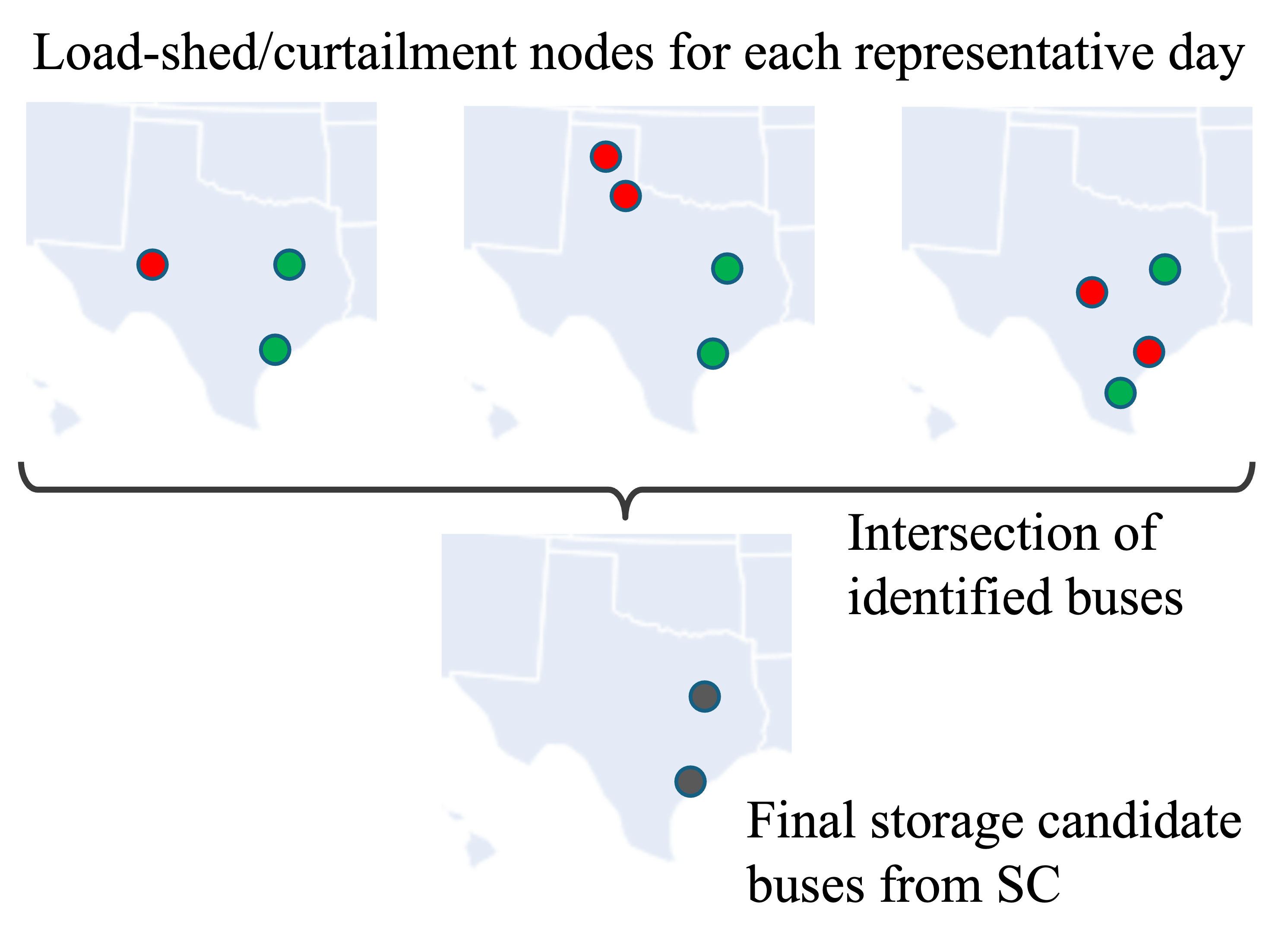}
    \caption{Search Space Reduction using the Proposed SC Method.}
    \label{fig:SCmethod}
\end{figure}

This paper employs a simple yet effective method for \emph{search space reduction}, identifying a subset of buses, referred to as \emph{Storage Candidates} (SC), for storage siting. The process begins by solving the recourse problem for each representative day without considering any line or storage investments. Nodes that experience load shedding or renewable curtailment during any hour are identified as SC. Preliminary experiments with the full TEP+Storage formulation revealed that these buses were typically selected for storage investment. This method aligns with existing practices, such as co-locating storage with renewable generation -- particularly solar installations -- or placing batteries in load centers to manage peak load hours. When multiple representative days are used, the intersection of nodes experiencing load shedding or curtailment across all days is considered. By considering only this reduced set of buses for storage siting, the TEP+Storage model becomes computationally tractable, even when solved at high spatial resolutions. Figure~\ref{fig:SCmethod} illustrates the SC method.

Table \ref{tab:candidates_computation} highlights the compute times for various configurations in 2030, comparing results with and without SC across different numbers of representative days, denoted by $k$. When considering all nodes for storage with a single representative day, the full model timed out after 72 hours. \emph{However, applying the SC heuristic made the problem computationally feasible.} Interestingly, the results suggest that the number of storage candidate nodes has a greater impact on computational challenge than the total number of binary variables. For instance, even with additional representative days (and more binary variables), the reduced number of storage candidate nodes led to lower computation times, as seen by the results for $k=5$ as compared to $k=1$.

    \begin{table}[!t]
        \centering
        \caption{Storage Candidates Computation Time}
        \label{tab:candidates_computation}
        \begin{tabular}{lcccc}
            \toprule
            \textbf{k} & \textbf{Candidates} & \textbf{Binary variables} & \textbf{Computation time} \\
            \midrule
            $k=1$ & 2000 & 50,000 & N/A (timed out) \\
            $k=1$ w. SC & 147 & 3,682 & 4.8 hours \\
            $k=5$ w. SC & 70 & 8,598 & 0.8 hours \\
            \bottomrule
        \end{tabular}
    \end{table}
    
\subsection{Modelling Investments}
Table \ref{tab:cost_config} provides details on investment costs and penalties for unserved and overserved energy. Storage investments decisions are restricted to sites identified by the SC method. The maximum power and energy ratings were set at 3 GW and 3 GWh, respectively. All storage devices were constrained to short-duration operation, with a maximum duration of 4 hours. Every system is assumed to have charge and discharge efficiency of 95\%, resulting in a 90.25\% roundtrip efficiency. 

Transmission line investments are offered as three upgrade options, corresponding to capacity increases of 30\%, 60\%, and 90\%. These reflect the enhancements achievable through typical line reconductoring processes \cite{GridLab2024Reconductoring}. All transmission lines in the network are candidates for capacity upgrades.

The capacity upgrade costs are adapted from \cite{Esmaili} while the storage installation costs are estimated using data from NREL’s Cost Projections for Utility-Scale Battery Storage \cite{NRELBatteryCost}. To prioritize reliability and adequacy, the selected penalty for load shedding reflects the high societal and economic costs of any load shed events.


    \begin{table}[!t]
        \centering
        \caption{Cost Configuration}
        \label{tab:cost_config}
        \begin{tabular}{cr}
        \toprule
         & Costs \\
        \midrule
        Unserved \& Overserved Energy          & \$2.5M/MWh       \\
            Capacity Upgrades &  \$1243/MW-km       \\ 
            Storage Fixed Cost           & \$500,000          \\ 
            Storage Energy Rating Cost          & \$120,000/MWh          \\ 
            Storage Power Rating Cost          & \$160,000/MW          \\ 
        \bottomrule
        \end{tabular}
    \end{table}

%% file: casestudy.tex
\section{Case study}
\label{sec:casestudy}

The paper presents numerical experiments based on the ACTIVSg2000 test case, a synthetic 2000-bus Texas power system developed by Texas A\&M University (TAMU). The system consists of 2000 buses, 3206 lines, 423 non-renewable generators, and 175 renewable generators which include solar, wind, offshore wind, and hydro. 

\paragraph{Network Data}
\label{sec:casestudy:data}
The synthetic ACTIVSg2000 test case is built from public information and statistical analysis of real power systems. These networks, while geographically unrelated to real grids and free of Critical Energy Infrastructure Information (CEII), provide a valuable platform for researchers to test and validate tools and techniques as if operating on real grids. The methodology to create such cases is detailed in \cite{synthetic_1}.



\begin{figure}[!t]
\centering
\includegraphics[width=0.95\columnwidth, trim=3cm 2cm 3cm 2cm, clip]{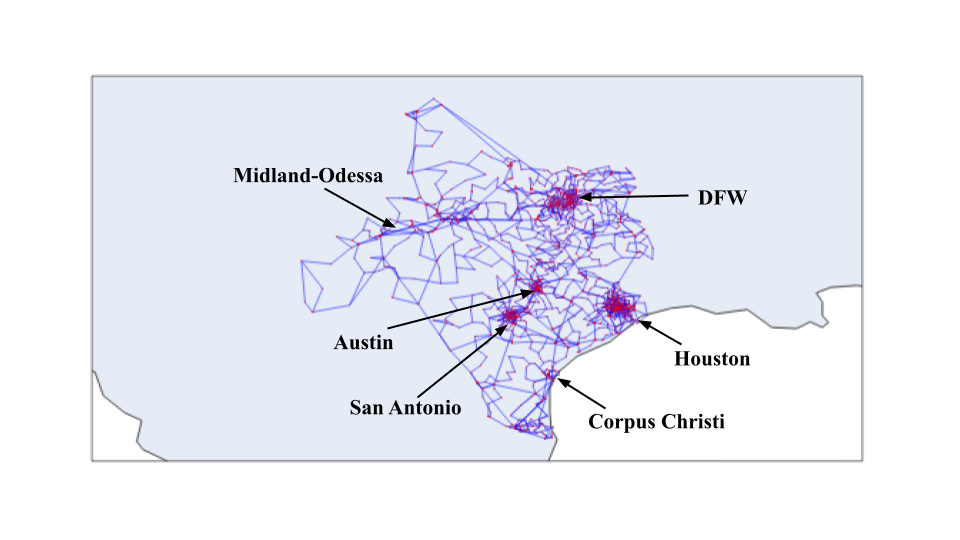}
\caption{Topology of Texas grid, with notable load centers labelled. Blue lines indicate branches, and red nodes indicate busses.}
\label{fig:texas_labelled}
\end{figure}

\begin{figure}[!t]
\centering
\includegraphics[width=0.99\columnwidth]{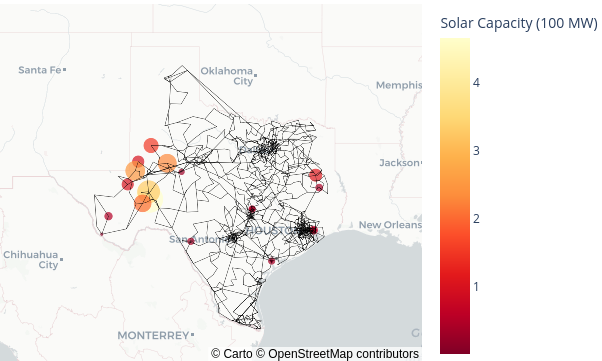}
\caption{Solar capacity of the system. Circles indicate the location of generators, with color and size corresponding to capacity in MW.}
\label{fig:solar_cap}
\end{figure}

\begin{figure}[!t]
\centering
\includegraphics[width=0.99\columnwidth]{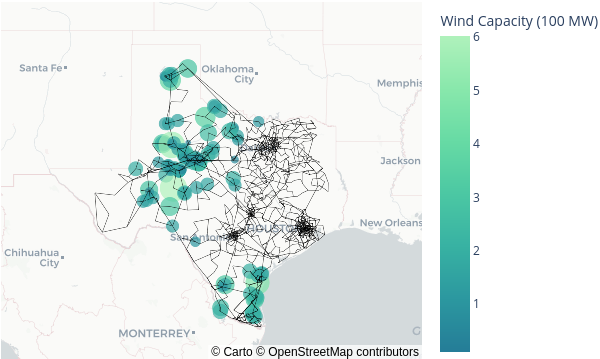}
\caption{Wind capacity of the system. Circles indicate the location of generators, with color and size corresponding to capacity in MW.}
\label{fig:wind_cap}
\end{figure}

\begin{figure}[!t]
\centering
\includegraphics[width=0.99\columnwidth]{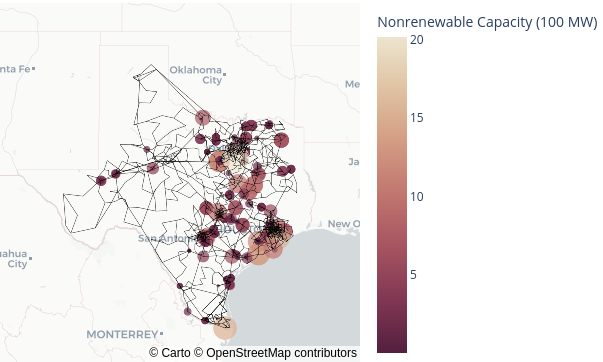}
\caption{Nonrenewable capacity of the system. Circles indicate the location of generators, with color and size corresponding to capacity in MW.}
\label{fig:nonrenewable_cap}
\end{figure}

Figures \ref{fig:solar_cap}, \ref{fig:wind_cap}, and \ref{fig:nonrenewable_cap} illustrate the geographic distribution of solar, wind, and nonrenewable capacity within the Texas system, overlaid on the grid topology. As depicted in Figure \ref{fig:texas_labelled}, the Texas system features several major metropolitan and industrial load centers, particularly Fort Worth-Dallas (northeast), Houston (east), and San Antonio and Austin (central), which are characterized by a higher density of nodes. Nonrenewable capacity is predominantly located in the eastern region, closer to these load centers and the oil and gas operations along the Gulf Coast. In contrast, solar and wind capacities are largely concentrated in the western region.

Within the western region, the Midland-Odessa area, situated in the Permian Basin, is notable for its generation mix, primarily driven by wind capacity with some nonrenewable generation present. The load in this area is largely industrial, supporting oil and gas activities. Further west, significant solar capacity is located, while wind capacity is more broadly distributed beyond this load center. These regional characteristics shape the results presented in this paper, which are dependent on the assumed generation configurations and may vary in other settings.

\paragraph{Future Generation Mix}

The shift in the generation mix toward renewable energy from 2030 to 2050 is modeled using projections from the U.S. Energy Information Administration (EIA) Annual Energy Outlook 2023 \cite{EIA2023Outlook}. Solar and wind generator capacities in the test case were uniformly scaled to reflect the projected growth for each type. Similarly, the capacities of coal, natural gas, and nuclear generators were adjusted based on the projected changes for these sources. Table \ref{tab:gen_load_scale} summarizes the scaled percentages for each generation type and load relative to the base year, 2022.

\begin{table}[!t]
        \centering
        \caption{Generation and Load Scaling}
        \label{tab:gen_load_scale}
        \begin{tabular}{lcccccc}
            \toprule
            \textbf{Type} & \textbf{2022} & \textbf{2030} & \textbf{2035} & \textbf{2040} & \textbf{2045} & \textbf{2050} \\
            \midrule
            Coal & 1.00 & 0.82 & 0.82 & 0.82 & 0.82 & 0.82 \\
            Natural Gas & 1.00 & 0.79 & 0.73 & 0.71 & 0.72 & 0.72 \\
            Nuclear & 1.00 & 0.98 & 0.90 & 0.80 & 0.80 & 0.80 \\
            Solar & 1.00 & 4.51 & 6.00 & 6.87 & 8.04 & 9.26 \\
            Wind & 1.00 & 2.02 & 2.23 & 2.26 & 2.32 & 2.43 \\
            Load & 1.00 & 1.13 & 1.21 & 1.31 & 1.41 & 1.52 \\
            \bottomrule
        \end{tabular}
    \end{table}
    
\paragraph{Load Projections}

Load growth is modeled with a year-to-year increase of 1.5\%  and was scaled uniformly across all buses. Higher growth rates were also tested, but these led to resource adequacy issues as the system approached 2050. Addressing such scenarios would require more detailed capacity expansion inputs, which the model is equipped to accommodate.

\paragraph{Representative Days}
To model operations throughout the year, five representative days are selected using k-medoids clustering based on the load, wind, and solar time series (i.e. $k=5$). Each day is weighted according to the size of its respective cluster. The average values for load, wind, and solar across the selected representative days are summarized in Table \ref{tab:rep_days}. 

    \begin{table}[!t]
        \centering
        \caption{Representative Day Averages in 2022}
        \label{tab:rep_days}
        \begin{tabular}{lccccc}
            \toprule
             & 01-14 & 05-15 & 07-27 & 09-09 & 10-21 \\
            \midrule
            Avg. Load (GW) & 35.1 & 32.5 & 48.0 & 49.2 & 33.7  \\
            Avg. Wind (GW) & 8.9 & 4.1 & 3.4 & 8.2 & 4.9  \\
            Avg. Solar (GW) & 0.8 & 0.5 & 0.8 & 0.8 & 0.9  \\
            \bottomrule
        \end{tabular}
    \end{table}

\paragraph{Planning Timeline}
The TEP+Storage model is implemented on the Texas system for 2030 to 2050, with investment decisions evaluated at 5-year intervals. To ensure continuity, investments made in earlier years are carried forward and incorporated into the optimization for subsequent years, resulting in cumulative investment costs. At each investment stage, the SC methodology is applied. For instance, identifying storage candidates for 2035 involves solving the recourse problem to identify nodes with load-shedding or renewable curtailment, while also considering the investments made in 2030.

\subsection{Computational Specifications}
\label{sec:casestudy:computational_specs}

The TEP+Storage model is developed in Julia utilizing the JuMP modeling language \cite{Lubin2023}. All experiments are conducted on 24-core Intel Xeon machines with Linux, hosted on the Phoenix cluster at GeorgiaTech \cite{PACE}. The experiments are run with 24 threads, 288 GB of RAM, a 72-hour time limit, and an optimality gap tolerance of 1.0\%, while other settings remain at their default values.

%% file: results.tex
\section{Experimental Results}
\label{sec:results}

\begin{table*}[!t]
    \caption{Investment decisions and costs for the three investment configurations: TEP+Storage, Transmission only, and Storage only. Results not applicable to the configuration are indicated as such with a dash, `-'.}
    \centering
    \label{tab:investments-100}
        \begin{tabular}{clcccccccccc}
        \toprule
         & \multirow{2}{*}{Year} & \multirow{2}{*}{\# Lines} & \multirow{2}{*}{\begin{tabular}[c]{@{}c@{}}\# Storage units / \\ Total capacity (GWh)\end{tabular}} & \multirow{2}{*}{SC} & \multicolumn{2}{c}{\# max investments} & \multicolumn{2}{c}{$CapEx$} & \multirow{2}{*}{$GenEx$ (\$B)} & \multirow{2}{*}{Load shed (GWh)} &  \\
         & &  &  &  & Lines & Storage & Lines (\$M) & Storage (\$B) &  &  &  \\
         \midrule 
         \parbox[t]{2mm}{\multirow{5}{*}{\rotatebox[origin=c]{90}{TEP+Storage}}} & 2030 & 100 & 0 / 0 & 61 & 12 & 0 & 50.18 & -- & 9.704 & 0 \\
        & 2035 & 134 & 0 / 0 & 54 & 20 & 0 & 87.44 & -- & 10.17 & 0 \\
        & 2040 & 181 & 3 / 0.54 & 52 & 34 & 0 & 146.1 & 0.149 & 10.83 & 0 \\
        & 2045 & 257 & 20 / 26.48 & 53 & 55 & 4 & 241.7 & 5.105 & 11.55 &  0 \\
        & 2050 & 392 & 51 / 109.6 & 43 & 122 & 28 & 481.2 & 21.37 & 12.11 & 3343\\
        \midrule 
        \parbox[t]{2mm}{\multirow{5}{*}{\rotatebox[origin=c]{90}{TEP Only}}} & 2030 & 100 & -- & -- & 12 & -- & 50.18 & -- & 9.704 & 0 \\
        & 2035 & 134 & -- & -- & 20 & -- & 87.44 & -- & 10.17 & 0 \\
        & 2040 & 179 & -- & -- & 40 & -- & 157.1 & -- & 10.84 & 15.2 \\
        & 2045 & 242 & -- & -- & 54 & -- & 282.1 & -- & 11.52 &  1759.6 \\
        & 2050 & 368 & -- & -- & 81 & -- & 447.8 & -- & 12.15 & 7200.0 \\
        \midrule
        \parbox[t]{2mm}{\multirow{5}{*}{\rotatebox[origin=c]{90}{Storage Only}}} & &  &  &  &  &  &  &  &  &   \\
        & 2030 & -- & 0 / 0 & 61 & -- & 0 & -- & 0 & 9.858 & 0 \\
        & 2035 & -- & 1 / 0.08690 & 92 & -- & 0 & -- & 0.01829 & 10.43 & 0 \\
        & 2040 & -- & 33 / 14.90 & 84 & -- & 1 & -- & 3.056 & 11.22 & 9.285 \\
        &  &  &  &  & &  & &  &  &   \\
        \bottomrule
        \end{tabular}
    \end{table*}
    
\begin{figure*}[p]
    \centering
    \subfloat[2030 TEP+Storage]{\includegraphics[width=0.275\textwidth, trim=5cm 2cm 0.8cm 2cm, clip]{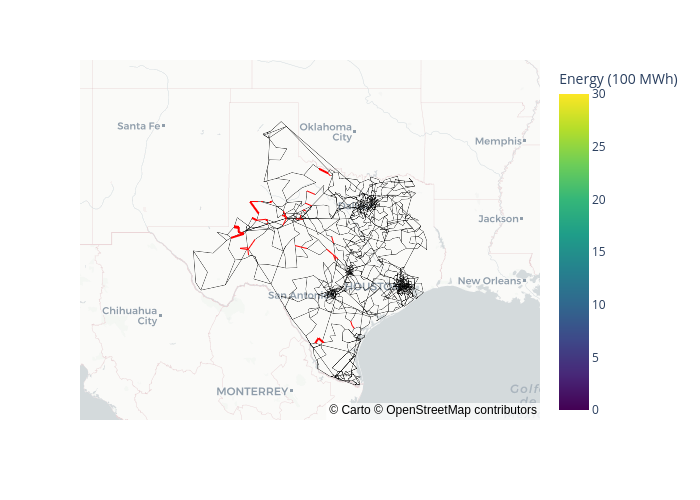} \label{fig:2030}} 
    \hspace{0.2cm}
    \subfloat[2030 TEP]{\includegraphics[width=0.275\textwidth, trim=5cm 2cm 0.8cm 2cm, clip]{fig/upgrades_2030.png} \label{fig:2030_onlyt}} 
    \hspace{0.2cm}
    \subfloat[2030 Storage]{\includegraphics[width=0.275\textwidth, trim=5cm 2cm 0.8cm 2cm, clip]{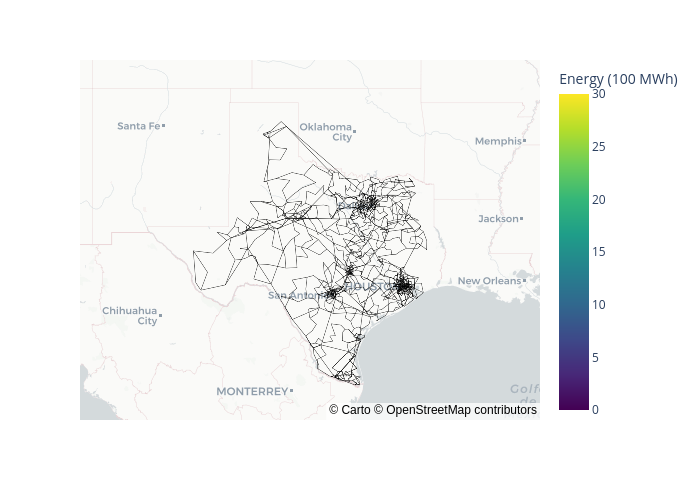} \label{fig:2030_onlys}}
    \\
    \subfloat[2035 TEP+Storage]{\includegraphics[width=0.275\textwidth, trim=5cm 2cm 0.8cm 2cm, clip]{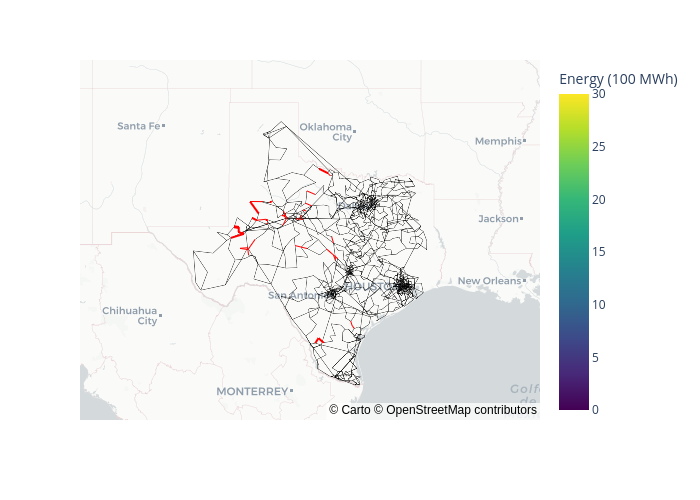} \label{fig:2035}} 
    \hspace{0.2cm}
    \subfloat[2035 TEP]{\includegraphics[width=0.275\textwidth, trim=5cm 2cm 0.8cm 2cm, clip]{fig/upgrades_2035.png} \label{fig:2035_onlyt}} 
    \hspace{0.2cm}
    \subfloat[2035 Storage]{\includegraphics[width=0.275\textwidth, trim=5cm 2cm 0.8cm 2cm, clip]{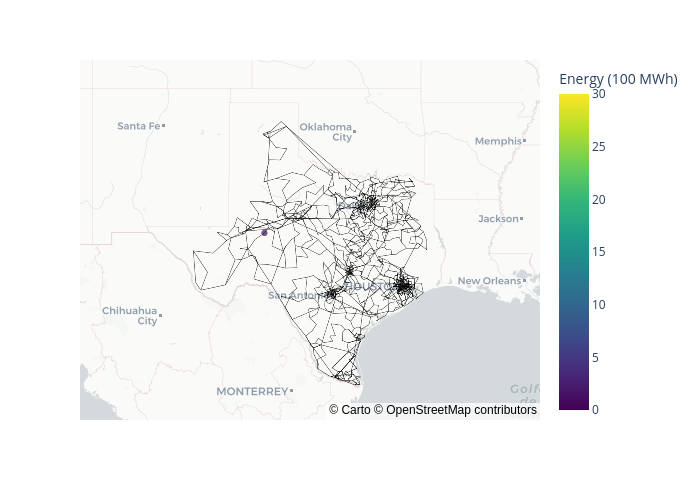} \label{fig:2035_onlys}}
    \\
    \subfloat[2040 TEP+Storage]{\includegraphics[width=0.275\textwidth, trim=5cm 2cm 0.8cm 2cm, clip]{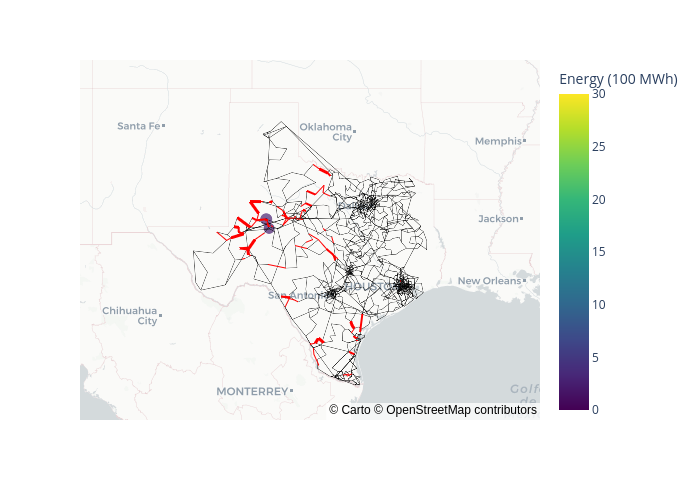} \label{fig:2040}} 
    \hspace{0.2cm}
    \subfloat[2040 TEP]{\includegraphics[width=0.275\textwidth, trim=5cm 2cm 0.8cm 2cm, clip]{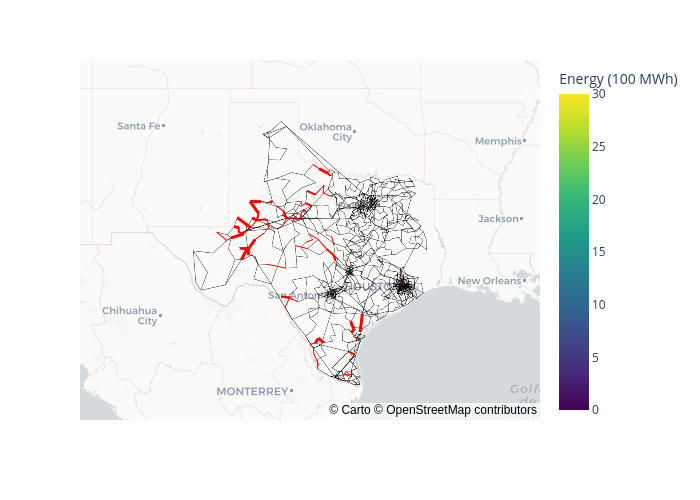} \label{fig:2040_onlyt}} 
    \hspace{0.2cm}
    \subfloat[2040 Storage]{\includegraphics[width=0.275\textwidth, trim=5cm 2cm 0.8cm 2cm, clip]{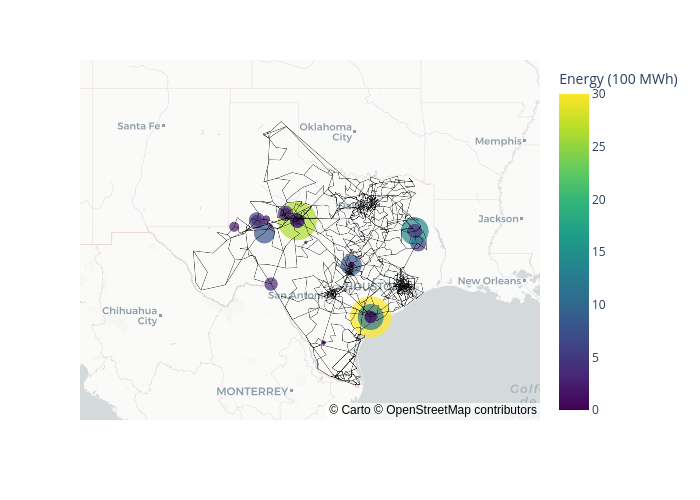} \label{fig:2040_onlys}}
    \\
    \subfloat[2045 TEP+Storage]{\includegraphics[width=0.275\textwidth, trim=5cm 2cm 0.8cm 2cm, clip]{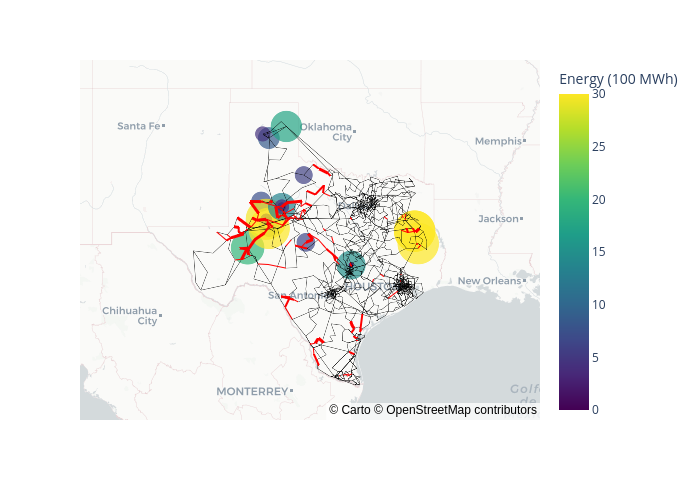} \label{fig:2045}} 
    \hspace{0.2cm}
    \subfloat[2045 TEP]{\includegraphics[width=0.275\textwidth, trim=5cm 2cm 0.8cm 2cm, clip]{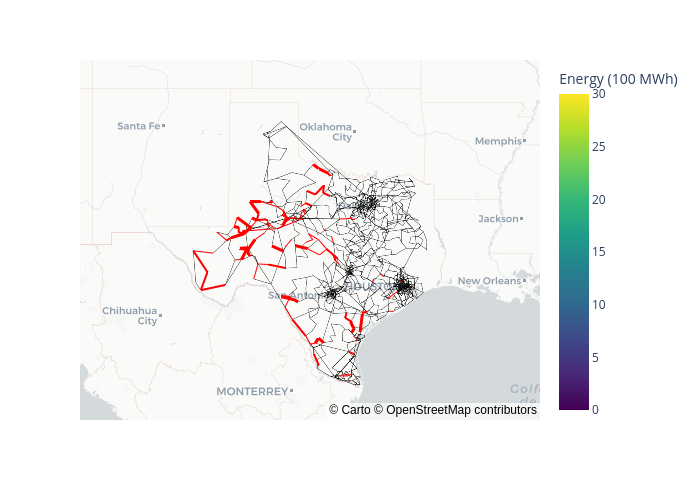} \label{fig:2045_onlyt}} 
    \hspace{0.2cm}
    \subfloat{\includegraphics[width=0.275\textwidth, trim=5cm 2cm 0.8cm 2cm, clip]{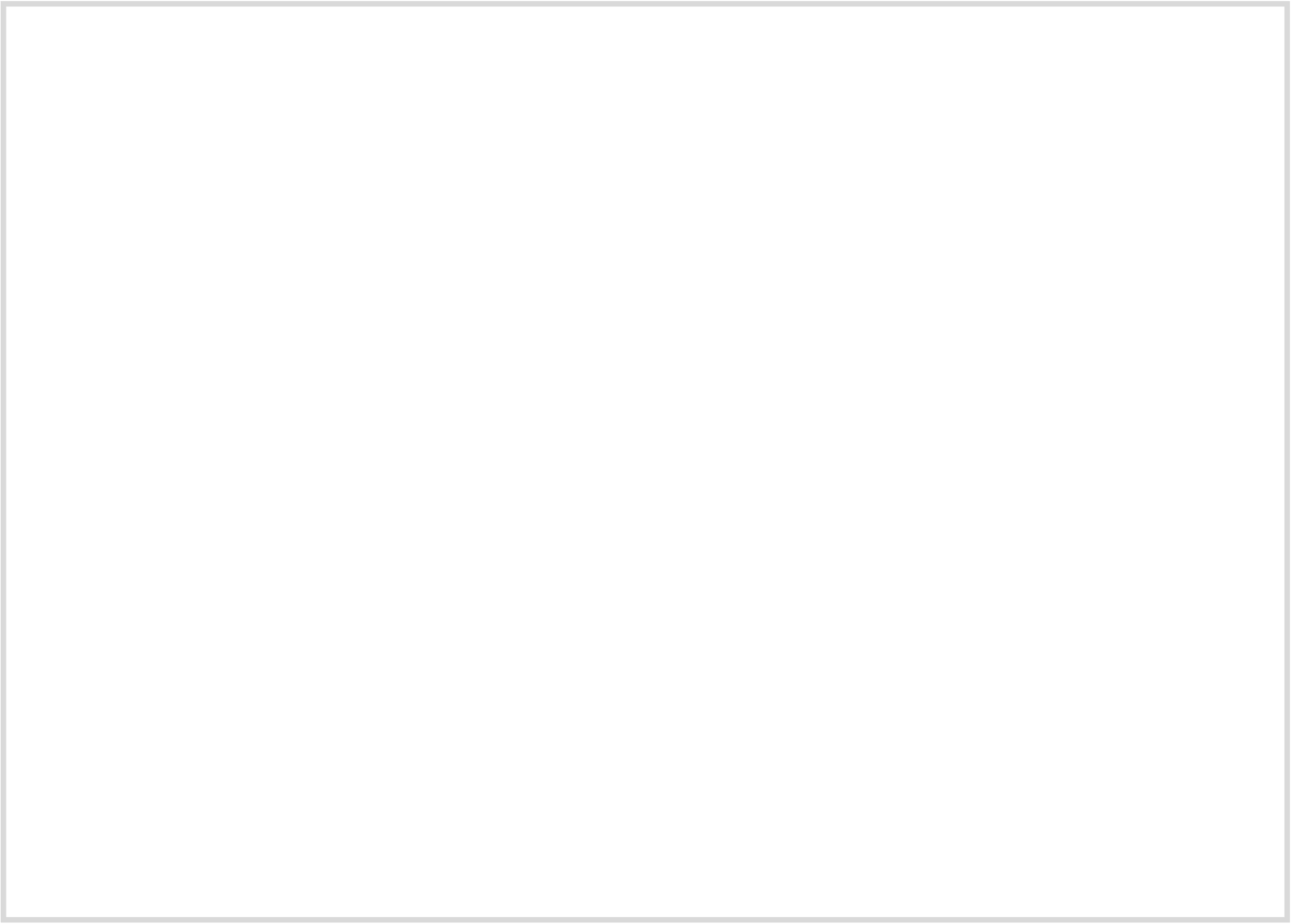} \label{fig:2045_onlys}}
    \\
    \subfloat[2050 TEP+Storage]{\includegraphics[width=0.275\textwidth, trim=5cm 2cm 0.8cm 2cm, clip]{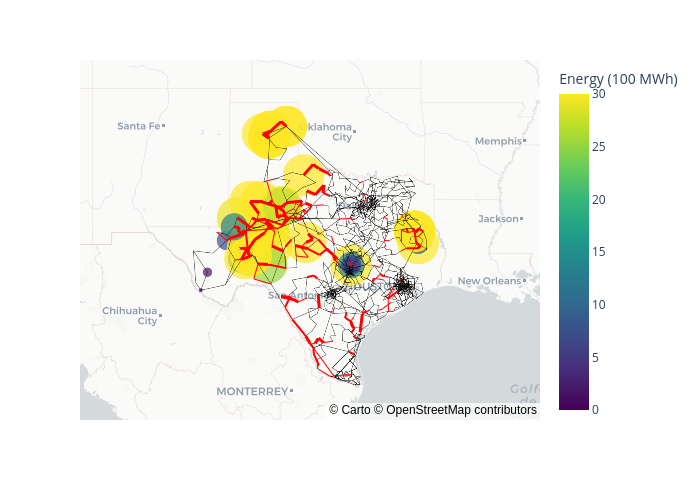} \label{fig:2050}} 
    \hspace{0.2cm}
    \subfloat[2050 TEP]{\includegraphics[width=0.275\textwidth, trim=5cm 2cm 0.8cm 2cm, clip]{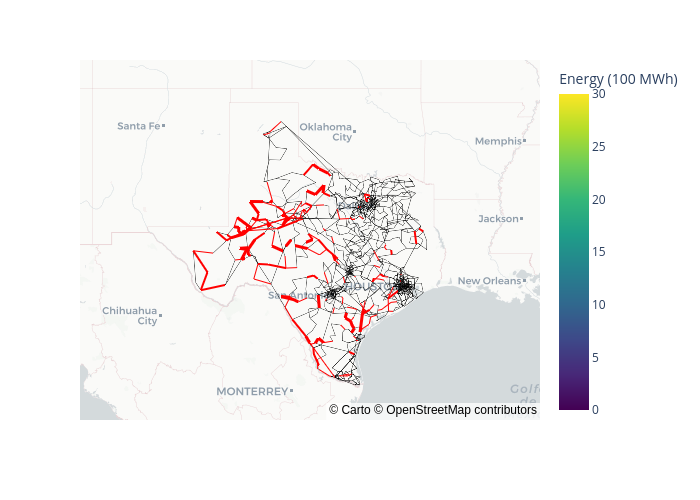} \label{fig:2050_onlyt}} 
    \hspace{0.2cm}
    \subfloat{\includegraphics[width=0.275\textwidth, trim=5cm 2cm 0.8cm 2cm, clip]{fig/upgrades_2045_onlys_placeholder.png} \label{fig:2050_onlys}}
    \caption{Upgrades from 2040 to 2050 in 5-year intervals, for all three configurations: \textbf{TEP+Storage} in the left column, \textbf{Transmission Only} in the middle column, and \textbf{Storage Only} in the right column. \textcolor{red}{\textbf{Red}} lines denote the locations of capacity upgrades, with their thickness corresponding to one of the three possible upgrade levels. Circles indicate the locations of storage, with their size and color representing the installed storage capacity in MWh.}
    \label{fig:upgrade_locations}
\end{figure*}   

\begin{figure}[!t] 
    \centering
    \subfloat[TEP+Storage in 2045]{%
        \includegraphics[width=0.80\columnwidth, trim=1cm 0 1cm 42, clip]{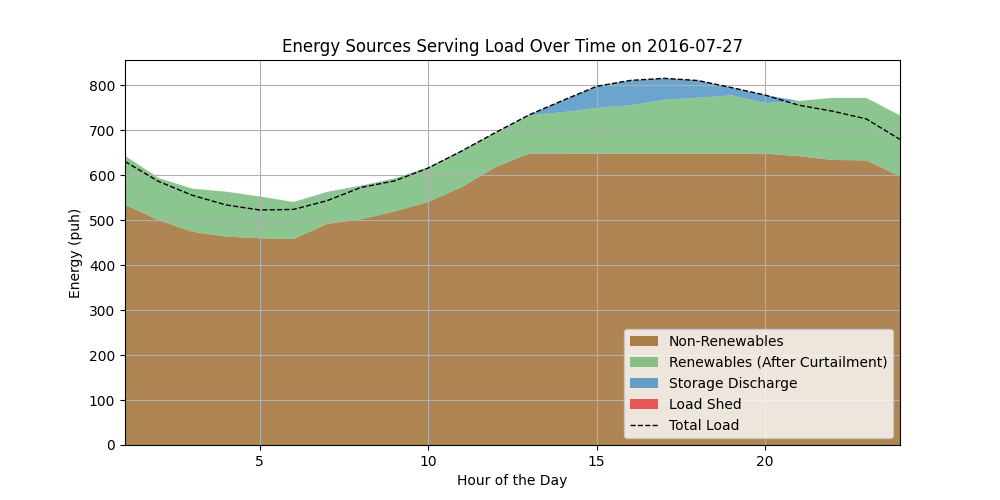} 
        \label{fig:serving_load_2045}
    }\\ 
    \subfloat[Transmission Only in 2045]{%
        \includegraphics[width=0.80\columnwidth, trim=1cm 0 1cm 42, clip]{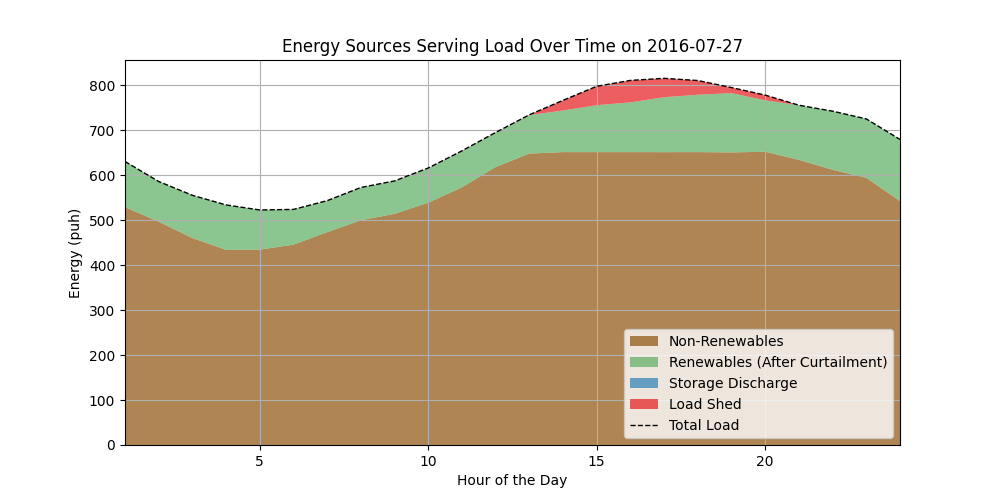} 
        \label{fig:serving_load_2045_onlyt}
    } \\
    \subfloat[TEP+Storage in 2050]{
        \includegraphics[width=0.80\columnwidth, trim=1cm 0 1cm 42, clip]{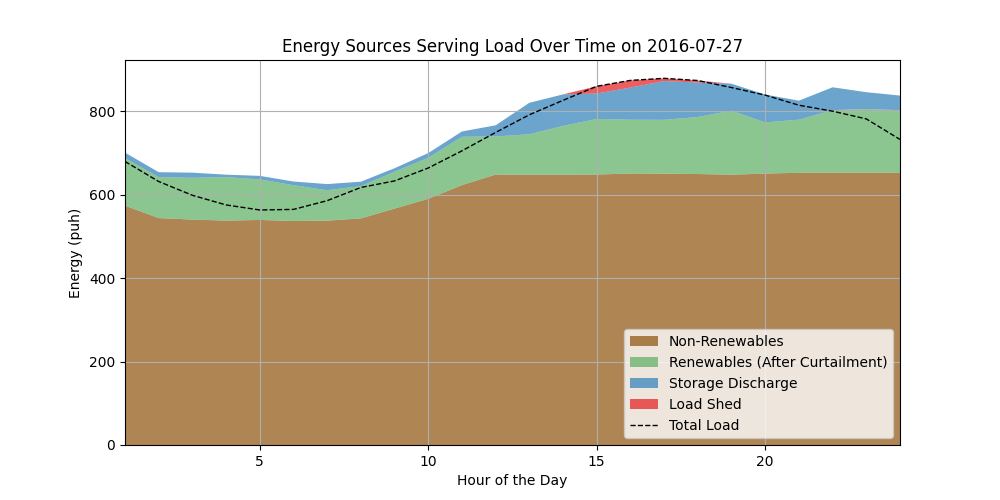} \label{fig:serving_load_2050}
        
    }
    \caption{Stacked plots showing the generation mix serving load on July 27 under the \textbf{TEP+Storage} and \textbf{Transmission Only} configurations. The plots account for the dispatch decisions of nonrenewable generators and any curtailment of renewable generation. The dashed line represents the load curve, with any area above it indicating storage charging. Units are in per-unit-hour (puh), where 1 puh corresponds to 100 MWh.}
    \label{fig:serving_load_comparison}
\end{figure}

\begin{figure*}[!t] 
    \centering
    \subfloat[Transmission Only on 07-27-2045 (Hours 14-20).]{%
        \includegraphics[width=0.3\textwidth, trim=5cm 2cm 0.8cm 2cm, clip]{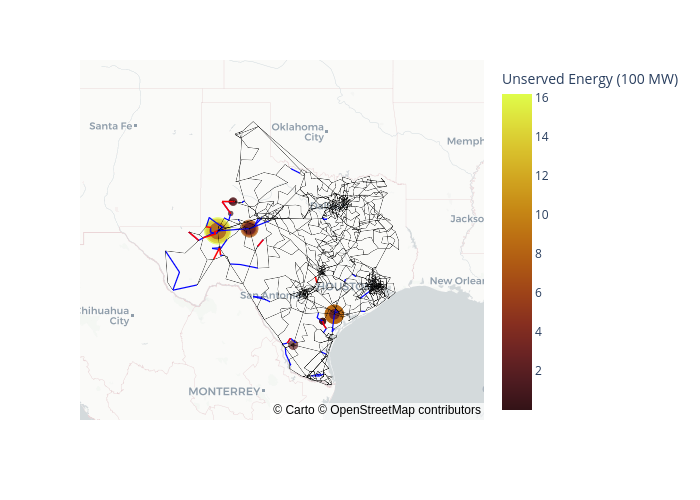} 
        \label{fig:2045_ue_onlyt}
    }
    \hfill
    \subfloat[TEP+Storage on 07-27-2050 (Hours 14-18).]{%
        \includegraphics[width=0.3\textwidth, trim=5cm 2cm 0.8cm 2cm, clip]{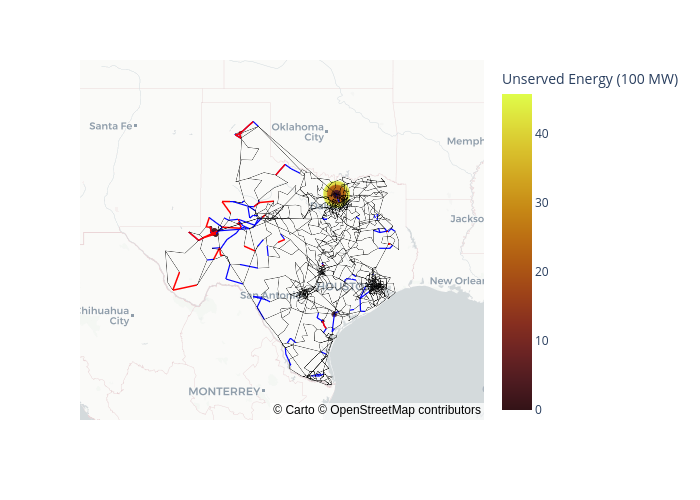} 
        \label{fig:2050_ue}
    }
    \hfill
    \subfloat[Storage Only on 09-09-2040 (Hours 14-16).]{%
        \includegraphics[width=0.3\textwidth, trim=5cm 2cm 0.8cm 2cm, clip]{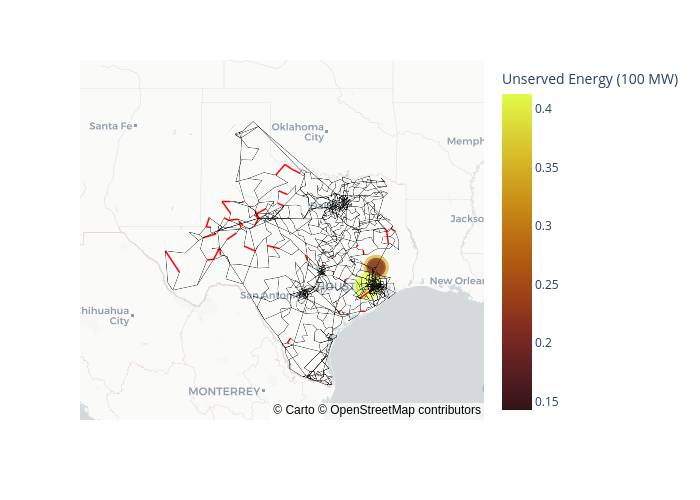} 
        \label{fig:2040_ue_onlys}
    }
    \vspace{0.15cm} 
    \subfloat[Transmission Only on 07-27-2045 (Hours 14-20).]{%
        \includegraphics[width=0.3\textwidth, trim=5cm 2cm 0.8cm 2cm, clip]{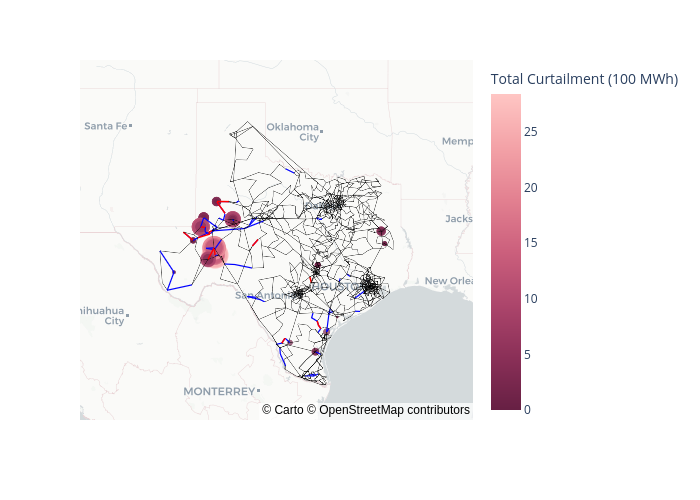} 
        \label{fig:2045_curtail_onlyt}
    }
    \hfill
    \subfloat[TEP+Storage on 07-27-2050 (Hours 14-18).]{%
        \includegraphics[width=0.3\textwidth, trim=5cm 2cm 0.8cm 2cm, clip]{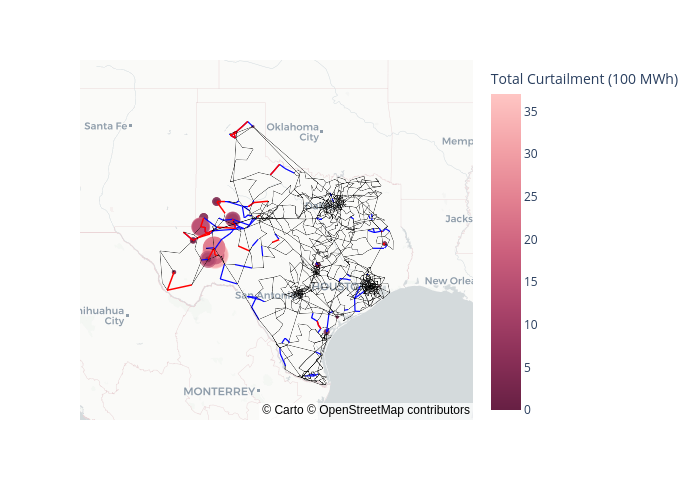} 
        \label{fig:2050_curtail}
    }
    \hfill
    \subfloat[Storage Only on 09-09-2040 (Hours 14-16).]{%
        \includegraphics[width=0.3\textwidth, trim=5cm 2cm 0.8cm 2cm, clip]{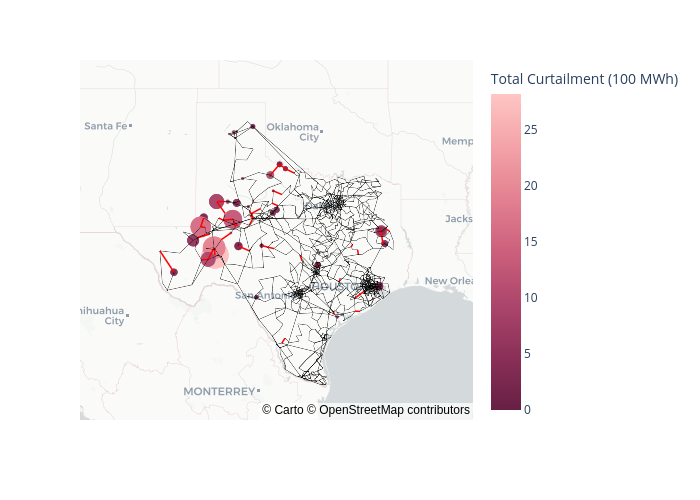} 
        \label{fig:2040_curtail_onlys}
    }
    \vspace{0.15cm} 
    \caption{Load shed events (top row) and curtailment (bottom row) for various configurations and time periods. \textcolor{blue}{\textbf{Blue}} lines represent branches where congestion was resolved through line upgrades, while \textcolor{red}{\textbf{red}} lines indicate branches that remain congested. Each column corresponds to a specific configuration: \textbf{Transmission Only} (left column, 07-27-2045), \textbf{TEP+Storage} (middle column, 07-27-2050), and \textbf{Storage Only} (right column, 09-09-2040).}
    \label{fig:operations_peakhours}
\end{figure*}

\begin{figure}[!t] 
    \centering
    \subfloat[Storage Only in 2040]{%
        \includegraphics[width=0.45\columnwidth, trim=5cm 2cm 7cm 3cm, clip]{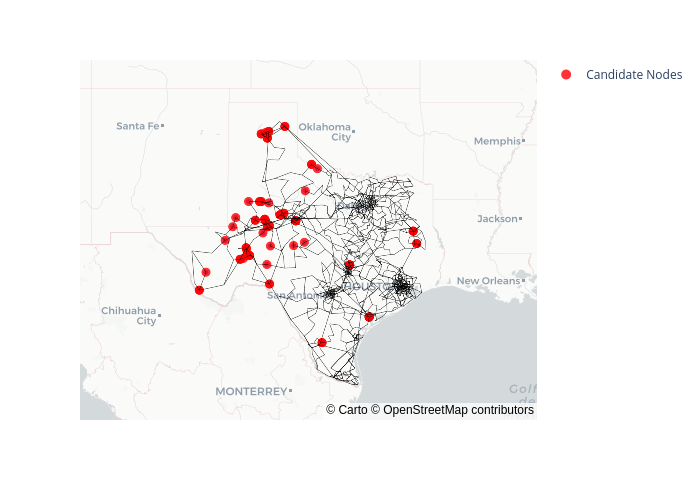}
        \label{fig:storage_candidates_2040_onlys}
    }
    \hspace{0.1cm}
    \subfloat[TEP+Storage in 2050]{%
        \includegraphics[width=0.45\columnwidth, trim=5cm 2cm 7cm 3cm, clip]{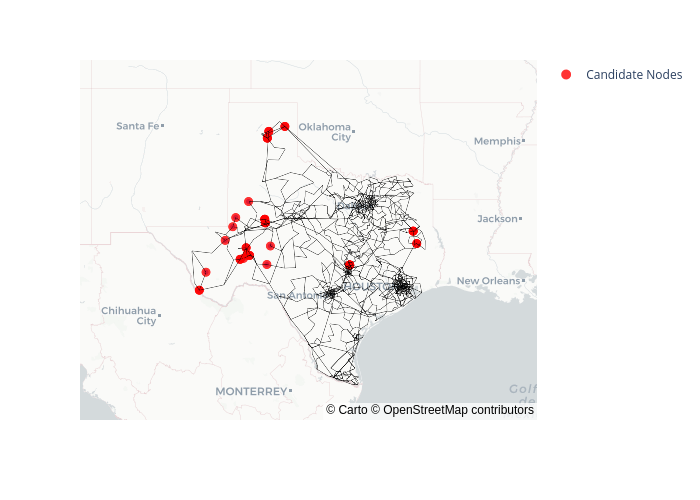}
        \label{fig:storage_candidates_2050}
    } 
    \caption{Map of candidate nodes for new storage devices using SC method, indicated by \textcolor{red}{\textbf{red}} circles.}
    \label{fig:storage_candidates}
\end{figure}

\begin{figure}[ht] 
    \centering
    \subfloat[Curtailment in 2040]{%
        \includegraphics[width=0.80 \columnwidth, trim=1cm 0 1cm 42, clip]{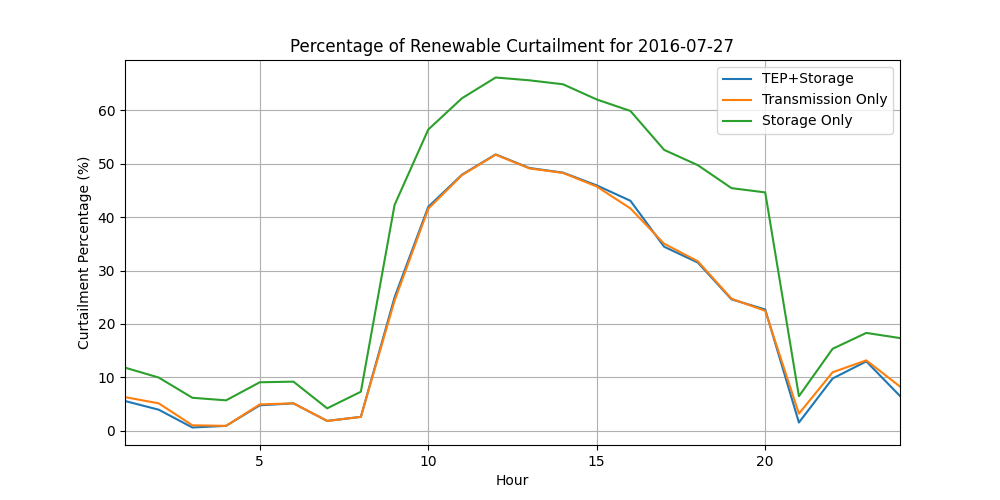} 
        \label{fig:curtailment_percentage_combined_2040}
    }\\ 
    \subfloat[Curtailment in 2050]{
        \includegraphics[width=0.80 \columnwidth, trim=1cm 0 1cm 42, clip]{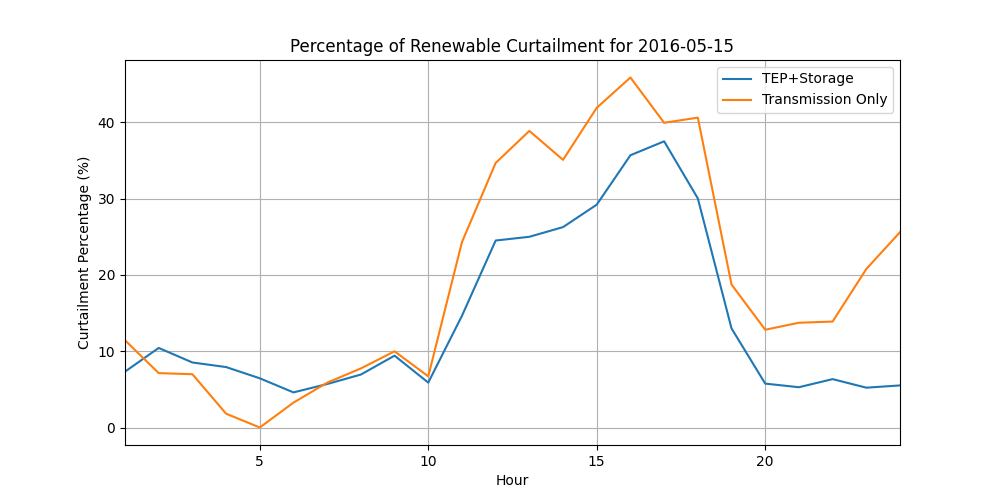} \label{fig:curtailment_percentage_combined_2050}
    }
    \caption{Curtailment as a percentage of renewable production on July 27 under the \textbf{TEP+Storage}, \textbf{Transmission Only}, and \textbf{Storage Only} configurations.}
    \label{fig:curtailment}
\end{figure}

The paper evaluates three investment configurations: TEP+Storage, Transmission Only, and Storage Only. Comparing the TEP+Storage model with the other two configurations highlights not only the individual contributions of transmission and storage technologies in supporting grid operations but also the synergy achieved when these technologies are co-optimized. The investment results and their associated costs are summarized in Table~\ref{tab:investments-100}. Figure~\ref{fig:upgrade_locations} plots the geographic locations of the resulting investment decisions across the three configurations. 

To evaluate system operations, the experiments consider the representative day characterized by high load and low wind generation (07-27), which presents a challenge in meeting load throughout the network. Figure~\ref{fig:serving_load_comparison} compares the generation mix serving load on this representative day for 2045 and 2050 over two configurations: TEP+Storage and Transmission Only. Figure~\ref{fig:operations_peakhours} plots the geographic distribution of renewable curtailment, load shed, and line congestion for different configurations during the peak load hours 14 to 20. 

It is important to note that this network model is synthetic, with both load and generation growth modeled using external data. As such, any results involving load-shedding events should not be interpreted as representative of future reliability concerns for the real Texas system.

\subsection{Key Observations}

The key results, which are further discussed in the next sections, can be summarized as follows. 
\begin{itemize}
    \item Co-investment in line upgrades and storage investments is necessary to meet electricity demand of future years while keeping costs low. Neither technology alone can meet future load.
    \item Transmission is the primary investment pathway to enable high utilization of zero-cost renewable energy by reducing congestion in critical regions.
    \item Storage is the primary investment pathway to satisfy peak demand in the afternoon via peak shaving and load shifting actions. 
    \item Storage remains an expensive technology, with delayed investments in favor of cheaper transmission upgrades. However, storage is necessary to provide temporal flexibility in generation.
\end{itemize}

\subsection{Spatial and Temporal Distribution of Investments}

Transmission lines are the key investment decisions in the TEP+Storage model, with the number of lines upgrading steadily increases at each 5-year interval. By 2050, almost 12\% of all transmission lines in the model are selected to be upgraded (392 lines), and 30\% of the selected lines are upgraded to the maximum level (i.e. a 90\% increase in capacity). These upgrades are concentrated primarily in the South and West, i.e., in regions with high solar and wind capacity. This trend is especially evident in the Midland-Odessa region, where numerous line upgrades are observed to facilitate the transfer of renewable energy from surrounding areas to serve local demand and to charge storage units in year 2040 and beyond. 

A similar trend for transmission upgrades is observed for the Transmission Only configuration, with minor differences in the number and level of line upgrades overall. \emph{Transmission upgrades are critical in regions experiencing significant renewable growth, while urban load centers with well-situated nonrenewable generation require comparatively less transmission investment.} 

In contrast, storage investments are observed much later, with the first units appearing in 2040, followed by considerable increase in installed capacity in 2045 and 2050. {\em Storage installations are concentrated in regions with significant solar capacity, as well as in major load centers.} These placements align with the paradigm set by the SC heuristic, targeting areas that may experience high volumes of curtailment and/or load shedding otherwise.

\subsection{Role of Transmission: Utilizing Renewable Generation}
In the Transmission Only case, load shedding appears as early as 2040, as compared to TEP+Storage where load shedding is delayed until 2050. Critically, the Transmission Only configuration is unable to meet the afternoon peak load.

Figures~\ref{fig:2045_ue_onlyt} and \ref{fig:2045_curtail_onlyt} plot the geographic distribution of load shed and renewable curtailment aggregated across the peak load hours (hours 14 to 20) in 2045, respectively. The results are overlaid on the transmission network, where upgraded lines (previously congested but resolved through upgrades) are shown in blue, and lines that remain congested are shown in red. The load shedding events are predominantly located in West Texas. Some load shedding occurs in urban load centers, such as Corpus Christi in the southeast. In this particular region, a congested line (which has been upgraded) limits the flow of additional renewable generation to the load center: load shedding occurs at the northern bus, while curtailment occurs at the Southern bus. 

This observation also holds in the Western part of the state, where both curtailment and load shedding occur due to line congestion despite significant upgrades in that region of the grid. Curtailment is heavily concentrated in the solar capacity hubs, further reinforcing the importance of significant transmission upgrades allocated to these regions. Figure~\ref{fig:curtailment} plots the renewable curtailment over 24-hours of the 07-27 representative day. 

The 2040 plot (Figure \ref{fig:curtailment_percentage_combined_2040}) clearly illustrates the relative importance of transmission upgrades compared to storage in increasing renewable utilization throughout the day. Without transmission upgrades, over 60\% of the available renewable generation is curtailed during the peak production hours, while with transmission upgrades to relieve congestion, this number drops to 50\%. 

\emph{In summary, transmission upgrades are critical for alleviating system congestion, especially in Western Texas. However, they are insufficient on their own to meet future load.}


\subsection{Role of Storage: Satisfying Load with Flexibility}

Battery storage in Eastern load centers is critical for providing flexibility in generation throughout the day, primarily for peak shaving applications. Storage units are typically charged during low load hours (early AM) and discharged during the peak hours of 14 to 20. During this time, the dispatchable nonrenewable generators are producing at maximum capacity, as seen by the flat dispatch line in Figure~\ref{fig:serving_load_comparison}, and congestion in the network limits the further use of renewable generation. 

Storage units located in the high renewable Western nodes play a critical role in increasing renewable utilization, and are located with a significant number of investments in the Western and Northern regions of the grid in 2045 and 2050. 

The co-investment of transmission upgrades and storage eliminates all load shedding events until 2050, while the Transmission Only and Storage Only cases both begin experiencing load shedding events a decade earlier in 2040. Figure~\ref{fig:curtailment} plots the renewable curtailment over 24-hours of the 07-27 representative day. While transmission is critical to reducing congestion and enabling renewable utilization in the first few investment periods (see 2040 plot), storage is critical in the last decade in order to supply peak load and further increase renewable utilization, as evident in the 2050 plot.

\emph{In summary, co-investment in line upgrades and storage investments is necessary to address regional challenges: meeting high demand in the Eastern load centers through the peak hours and mitigating renewable congestion issues in the West. Transmission upgrades in the early years enable effective storage utilization in future years.} 

\subsection{Trade-offs between Transmission and Storage}
To evaluate the impact of storage technology costs on investment decisions, a sensitivity analysis was performed. The experiments consider 75\%, 125\%, and 150\% of the baseline parameter cost levels for storage detailed in Table~\ref{tab:cost_config}. These represent potential cost reductions from technological improvements or policy actions, and potential cost increases from higher materials or manufacturing costs from critical metal supply chains or on-shoring of manufacturing. Table~\ref{tab:investments-storagecosts} shows results for different storage cost levels. 

\begin{table*}[!t]
    \caption{Investment decisions and costs for TEP+Storage at different configurations of storage cost levels. Results not applicable to the configuration are indicated as such with a dash, `-'.}
    \label{tab:investments-storagecosts}
        \begin{tabular}{clcccccccccc}
        \toprule
        & \multirow{2}{*}{Year} & \multirow{2}{*}{\# Lines} & \multirow{2}{*}{\begin{tabular}[c]{@{}c@{}}\# Storage units / \\ Total capacity (GWh)\end{tabular}} & \multirow{2}{*}{SC} & \multicolumn{2}{c}{\# max investments} & \multicolumn{2}{c}{$CapEx$} & \multirow{2}{*}{$GenEx$ (\$B)} & \multirow{2}{*}{Load shed (GWh)} &  \\
         &  &  &  &  & Lines & Storage & Lines (\$M) & Storage (\$B) &  &  &  \\
         \midrule 
        \parbox[t]{2mm}{\multirow{4}{*}{\rotatebox[origin=c]{90}{75\%}}} & 2030 & 100 & 0 / 0 & 61 & 12 & 0 & 50.18 & -- & 9.704 & 0 \\
        & 2035 & 134 & 0 / 0 & 54 & 20 & 0 & 87.44 & -- & 10.17 & 0 \\
        & 2040 & 180 & 4 / 0.5406 & 52 & 34 & 0 & 145.7 & 0.1134 & 10.83 & 0\\
        & 2045 & 256 & 20 / 26.52 & 53 & 58 & 4 & 238.6 & 3.836 & 11.55 & 0 \\
        \midrule 
        \parbox[t]{2mm}{\multirow{5}{*}{\rotatebox[origin=c]{90}{100\%}}} & 2030 & 100 & 0 / 0 & 61 & 12 & 0 & 50.18 & -- & 9.704 & 0 \\
        & 2035 & 134 & 0 / 0 & 54 & 20 & 0 & 87.44 & -- & 10.17 & 0 \\
        & 2040 & 181 & 3 / 0.5377 & 52 & 34 & 0 & 146.1 & 0.1492 & 10.83 & 0 \\
        & 2045 & 257 & 20 / 26.48 & 53 & 55 & 4 & 241.7 & 5.105 & 11.55 &  0 \\
        & 2050 & 392 & 51 / 109.6 & 43 & 122 & 28 & 481.2 & 21.37 & 12.11 & 3343\\
        \midrule
        \parbox[t]{2mm}{\multirow{4}{*}{\rotatebox[origin=c]{90}{125\%}}} & 2030 & 100 & 0 / 0 & 61 & 12 & 0 & 50.18 & -- & 9.704 & 0 \\
        & 2035 & 134 & 0 / 0 & 54 & 20 & 0 & 87.44 & -- & 10.17 & 0 \\
        & 2040 & 158 & 4 / 0.5967 & 52 & 34 & 0 & 121.7 & 0.2056 & 10.85 & 0 \\
        & 2045 & 258 & 26 / 26.48 & 57 & 59 & 5 & 244.2 & 6.311 & 11.55 &  0 \\
        \midrule
        \parbox[t]{2mm}{\multirow{3}{*}{\rotatebox[origin=c]{90}{150\%}}} & 2030 & 100 & 0 / 0 & 61 & 12 & 0 & 50.18 & -- & 9.704 & 0 \\
        & 2035 & 134 & 0 / 0 & 54 & 20 & 0 & 87.44 & -- & 10.17 & 0 \\
        & 2040 & 166 & 2 / 0.5378 & 52 & 31 & 0 & 129.4 & 0.2225 & 10.84 & 0\\
        \bottomrule
        \end{tabular}
    \end{table*}

The prevailing trends in transmission and storage investments remain the same, with storage investment delayed until 2040 even at reduced costs of 75\%. This limited sensitivity to cost reductions indicates that transmission is the dominant bottleneck in supplying loads at a low cost. However, in 2040 and 2045, comparable levels of storage investment are observed, even at higher cost levels of 125\% and 150\%. This limited sensitivity to cost increases indicates that storage is a necessary investment to meet future load, regardless of its cost. This is an expected result, given that the Transmission Only configuration struggles to meet load as early as 2040, and even the TEP+Storage configuration experiences some load shedding events in 2050. The storage investment decisions are more sensitive to the penalty for unserved energy. Lower penalties lead to reduced storage and capacity investments, favouring load-shedding as early as 2040.

\subsection{Impact of Modeling Decisions on Storage Investments}
A few comments must be made regarding the model and resulting investment decisions. First, the weighting of generation costs relative to storage costs in the objective function greatly influences the investment decisions, as seen from preliminary studies. The TEP+Storage model does not incorporate any financial incentives for storage investments and assumes that storage operations do not contribute to the $OpEx$. It also omits merchant-owned and operated storage, which is highly prevalent in the grid today \cite{Dvorkin,Sioshansi_2010_storageownershippricevolatility}.

Second, the hourly timescale used in the recourse problem does not capture the sub-hourly flexibility that storage can uniquely provide compared to other dispatchable generation resources. The increasing participation of storage in providing ancillary services highlights a need for considering sub-hourly operations \cite{kazemi2017operation}. It is expected that a TEP+Storage model with higher operational resolution will recommend additional storage investments to provide sub-hourly ramping needs. Other forms of operational flexibility can be used to support the grid; these include grid-enhancing technologies like dynamic line ratings and power flow control devices, demand response and flexible loads, and aggregations of distribution-level assets \cite{li2018grid}. 

Finally, in Model~\ref{model:TNEP}, the residual energy in the batteries at the end of the period (i.e., 24-hours) is constrained to be half the storage capacity, as per Eq.~\eqref{eq:TNEP:final_charge}. This operational requirement simplifies the use of representative days to evaluate performance over the year, by assuming batteries are always half full at the beginning of any representative period. However, inclusion of this constraint may limit the capabilities of the storage units: the load shedding observed in Fig.~\ref{fig:serving_load_2050} for TEP+Storage in 2050 could easily be resolved by discharging the battery units during the peak hours, at the expense of violating the residual energy constraint. This constraint is further studied in models optimizing battery operations over longer time periods using representative days and system states \cite{wang2022optimal, tejada2018enhanced}.

\subsection{Evaluating the SC method}
A limitation of the current storage candidates heuristic, while simple and effective in reducing the search space to maintain computational tractability, is its potential to overlook more efficient storage siting locations. This limitation is reflected in the persistence of load shedding observed in configurations of TEP+Storage in 2050, and Storage Only in 2040. Figure~\ref{fig:storage_candidates} shows the SC sites for these two configurations. With the TEP+Storage configuration, the Fort Worth-Dallas area experiences significant load shedding in 2050 (Figure \ref{fig:2050_ue}), but was not considered as a potential storage site for the TEP+Storage. This omission occurs because load shedding took place only on a subset of the representative days, not all days, leading to their exclusion from the candidate set. A similar observation is made for the Houston area with the Storage Only configuration in 2040 (Figure~\ref{fig:2040_ue_onlys}). 

Two additional experiments were run after observing the 2050 results. In the first experiment, storage was manually added to the Fort Worth-Dallas area at 3 nodes, at maximum power and energy ratings (3GW/3GWh batteries). The additional installed storage roughly reduced the total load shedding throughout the year by 25\%, eliminating load shed on day 09-09 which has high load and high wind and solar generation. However, the load shed on day 07-27 which has high load and low wind, still experiences load shedding. In the second experiment, nodes in the Fort Worth-Dallas area were added to the SC set and the investments are re-optimized. The model added storage to 3 nodes at maximum energy ratings (3GWh) and optimized operations to reduce annual load shed by 29\%. The storage additions from the TEP+Storage model did not have batteries at the maximum power rating, reflecting the need for capacity rather than high power discharge capabilities. 

These results indicate that an iterative approach is needed in the SC method, wherein the selected nodes are incremented in a data-driven fashion as revealed by the TEP+Storage investment decisions. Any load shedding centers identified after investments should be added to the SC set and the analysis repeated. However, the SC method must maintain computational tractability of the TEP+Storage problem. 

As discussed in Section \ref{sec:methodology}, modeling such a high-spatial-resolution case study poses significant challenges in maintaining computational tractability. This challenge was a key motivator for the development of the SC heuristic, which prioritizes the intersection of curtailment and load-shedding events across representative days, rather than their union, to limit the number of candidate nodes. However, as the number of storage candidates grows in later years, particularly in the Storage Only configuration, the SC heuristic struggles to maintain scalability. This is evidenced by timed-out instances in these configurations, where the increasing number of candidates leads to prolonged computation times. 

Additionally, compute times appear to be highly sensitive to cost configurations, particularly when the cost of storage investment increases and approaches the value of GenEx, as shown in Table \ref{tab:investments-storagecosts}. Incomplete rows in the table highlight instances where the model timed out after 72 hours on the high-performance computing (HPC) system. These observations underscore the fact that full scalability has not yet been achieved with the current methodology, leaving significant room for improving computational efficiency. Additional work is needed in identifying candidate solutions and accelerating the TEP+Storage solve time.

%% file: conclusion.tex
\section{Conclusion}
\label{sec:conclusion}

This paper presented a TEP+Storage formulation  for capacity expansion and applied it to a case study at a high spatial resolution of 2,000 buses. To achieve computational feasibility, the paper introduced a novel yet simple and effective method for reducing the search space of storage candidates. The case study evaluated both transmission and storage investments independently and jointly, providing insight into their individual contributions and synergistic impacts on satisfying future electricity demand. The results highlight geographic trends in transmission investments which are primarily driven by regions of renewable expansion, and storage investments which are influenced by renewable curtailment or load shedding. Storage is predominantly used to shave peak load hours and provide flexibility in meeting demand using generation from other time periods. {\em The findings emphasize that co-optimizing transmission and storage is crucial for minimizing costs and effectively meeting demand.}

The proposed SC method reduces the search space effectively,  and makes the TEP+Storage model computationally tractable. Its current implementation prioritizes reducing the search space, potentially at the expense of candidates which can further reduce cost or load shedding. The development of a robust SC method which prioritizes locations with persistent load shedding and evaluates the quality of the overall solution is the topic of future work. Future work will also consider high temporal resolutions to fully capture the value of storage in providing ancillary services, while maintaining a high spatial resolution. Additional techniques including search space reduction for transmission upgrades and new transmission lines, convex relaxations of the storage recourse problem, and acceleration algorithms like Benders decomposition will be explored to improve computational efficiency and scalability.